\documentclass[]{aa}  
\usepackage{graphicx}
\usepackage{txfonts}
\usepackage{supertabular}

\begin{document} 

\def\teff{\hbox{$T_{\rm eff}$}}  
\def\logg{\hbox{$\log g$}}  
\def\sn{\hbox{S/N}}  
\def\BmV{\hbox{$B-V$}}  
\def\vsin{\hbox{$v \sin i$}}  
\def\vrad{\hbox{$v_{\rm rad}$}}  
\def\vmic{\hbox{$\xi_{\rm mic}$}}  
\def\vmac{\hbox{$\xi_{\rm mac}$}}  
\def\kms{\hbox{km\,s$^{-1}$}}  
\def\ms{\hbox{m\,s$^{-1}$}}  
\def\mV{\hbox{$m_{\rm V}$}}   
\def\em{\it}  
\def\arcsec{\hbox{$^{\prime\prime}$}}  
\def\degr{\hbox{$^\circ$}}  
\def\rpd{\hbox{rad\,d$^{-1}$}}   
\def\omeq{\hbox{$\Omega_{\rm eq}$}}   
\def\peq{\hbox{P$_{\rm eq}$}}   
\def\dom{\hbox{$d\Omega$}}   
\def\kis{\hbox{$\chi^2$}}   
\def\kisr{\hbox{$\chi^2_{\rm r}$}}   
\def\drot{\hbox{differential rotation}}  
\def\epseri{\hbox{$\epsilon$~Eridani}}   
\def\halpha{\hbox{H$\alpha$}}   
\def\msun{\hbox{$M_\odot$}}   
\def\prot{\hbox{$P_{\rm rot}$}}

\title{Multi-instrumental view of magnetic fields and activity of $\epsilon$~Eridani with SPIRou, NARVAL, and TESS}
\titlerunning{Magnetism and activity of $\epsilon$~Eri}

\author{
P. Petit\inst{1}
\and
C.P. Folsom\inst{1,2}
\and
J.-F. Donati\inst{1}
\and
L. Yu\inst{1}
\and
J.-D. do Nascimento Jr.\inst{3,4}
\and
S. Jeffers\inst{5,6}
\and
S.C. Marsden\inst{7} 
\and
\\J. Morin\inst{8}
\and
A.A. Vidotto\inst{9}
}
         
\institute{
Institut de Recherche en Astrophysique et Plan\'etologie, Universit\'e de Toulouse, CNRS, CNES, 14 avenue Edouard Belin, 31400 Toulouse, France 
\and
{Department of Physics \& Space Science, Royal Military College of Canada, PO Box 17000 Station Forces, Kingston, ON, Canada K7K 0C6}
\and
{Departamento de F\'isica Te\'orica  e Experimental, Universidade Federal do Rio Grande do Norte, Natal, RN 59072-970, Brazil}
\and
{Harvard-Smithsonian Center for Astrophysics, 60 Garden St., Cambridge, MA 02138, USA}
\and
Universit\"at G\"ottingen, Institut f\"ur Astrophysik, Friedrich-Hund-Platz 1, 37077 G\"ottingen, Germany
\and
Max-Planck-Institut f\"ur Sonnensystemforschung, Justus-von-Liebig-Weg 3, 37077 G\"ottingen, Germany
\and
University of Southern Queensland, Centre for Astrophysics, Toowoomba, QLD 4350, Australia
\and
Laboratoire Univers et Particules de Montpellier, Universit\'e de Montpellier, CNRS, F-34095, France 
\and
School of Physics, Trinity College Dublin, University of Dublin, Dublin-2, Ireland
}
             
\date{}

\abstract
{}
{We report on observations of the active K2 dwarf \epseri\ based on contemporaneous SPIRou, NARVAL and TESS data obtained over two months in late 2018, when the activity of the star was reported to be in a non-cyclic phase.}
{Near-infrared spectropolarimetry was obtained using SPIRou over 4 nights in late September, while visible spectropolarimetry was collected with NARVAL over 20 nights, from 18 Sept to 07 Nov. We first recover the fundamental parameters of the target from both visible and nIR spectral fitting. The large-scale magnetic field is investigated from polarimetric data. From unpolarized spectra, we estimate the total magnetic flux through Zeeman broadening of magnetically sensitive nIR lines and the chromospheric emission using the CaII H\&K lines. The photometric monitoring, secured with TESS between 19 Oct and 15 Nov, is modelled with pseudo-periodic Gaussian Process Regression.}
{Fundamental parameters of \epseri\ derived from visible and near-infrared wavelengths provide us with consistent results, also in agreement with published values. We report a progressive increase of macroturbulence towards larger nIR wavelengths. Zeeman broadening of individual lines highlights an unsigned surface magnetic field $B_{\rm mono} = 1.90 \pm 0.13$ kG, with a filling factor $f = 12.5 \pm 1.7$\% (unsigned magnetic flux $Bf = 237 \pm 36$~G). The large-scale magnetic field geometry, chromospheric emission and broadband photometry display clear signs of non-rotational evolution over the course of data collection. Characteristic decay times deduced from the light curve and longitudinal field measurements fall in the range 30-40~d, while the characteristic timescale of surface differential rotation, as derived through the evolution of the magnetic geometry, is equal to $57 \pm 5$~d. The large-scale magnetic field exhibits a combination of properties not observed previously for \epseri, with a surface field among the weakest previously reported, but also mostly axisymmetric, and dominated by a toroidal component.}
{}


\maketitle

\section{Introduction}

Most Sun-like stars experience a strong magnetic activity during the first Gyr of their evolution, as a consequence of an efficient global dynamo triggered by their high spin rate. The variable phenomena induced at the stellar surface and in upper atmospheric layers span a wide range of spatial scales, temporal scales, and spread their signatures over most of the electromagnetic spectrum, shaping the extended environment of the star where young planets may orbit \citep{2016ApJ...820L..15D}. Widely studied manifestations of stellar activity include core emission in chromospheric lines \citep{1984ApJ...279..763N,2018A&A...616A.108B}, spectroscopic and broad band photometric signatures of dark spots and plages (see \citealt{2005LRSP....2....8B}, for a review), polarized Zeeeman signatures (e.g. \citealt{1990A&A...232L...1D,2014MNRAS.444.3517M}) and Zeeman broadening of magnetically-sensitive spectral lines \citep{Saar1988,2019A&A...630A..99L}. Owing to the diverse instrumentation and modelling tools needed to study these fragmented diagnoses, most studies concentrate on a specific tracer of activity, getting in return a very incomplete view of a complex phenomenon. Here, we propose to grasp a more diverse view of stellar activity for the specific case of \epseri{}, taking advantage of contemporaneous observations gathered with three different instruments.

With a distance of $3.2028 \pm 0.0047$~pc \citep{2018A&A...616A...1G}, \epseri\ is one of the closest solar-type stars. Being in the solar neighborhood, it benefited from interferometric radius measurements ($R = 0.74 \pm 0.01 R_\odot$, \citealt{diFolco2007,2012ApJ...744..138B}), confirming its status as a main sequence dwarf. From the combined knowledge of the distance and radius, it is possible to infer its surface black body temperature (\teff $=5076 \pm 30$ K, \citealt{heiter2015}). This estimate is consistent  with its K2V spectral type \citep{1989ApJS...71..245K} and mostly agrees with its spectroscopic effective temperature ($T_{\rm eff} = 5146 \pm 30$~K, \citealt{2005ApJS..159..141V}). Spectroscopic and interferometric studies also agree with each other on a mass estimate, with $0.856 \pm 0.08 M_\odot$ given by \cite{2005ApJS..159..141V} or $0.80 \pm 0.06 M_\odot$ from \citet{heiter2015}. 

Radio observations have unveiled a debris disc \citep{1986ASSL..124...61G} taking the form of a narrow ring with arc-like azimuthal structures \citep{2017MNRAS.469.3200B}. This circumstellar residual from the stellar formation is indicative of a young age ($439 \pm 52$~Myr was proposed by \citealt{2007ApJ...669.1167B}), and the geometrical ring properties may reveal a resonant interaction with massive planetary companions. A first planet was actually detected at 3.4~AU by \cite{2000ApJ...544L.145H} with a mass $M.\sin i = 0.86~M_{\rm J}$, and later confirmed by \cite{2012ApJS..200...15A}. This eccentric planetary companion ($e = 0.6$) may share the neighborhood of \epseri\ with a second, unconfirmed planet of $\approx 0.1~M_{\rm J}$ at 40 AU from the star, according to numerical simulations of the debris disc \citep{2002ApJ...578L.149Q}.

Due to its young age, \epseri\ is rotating relatively fast, with sustained CaII H\&K emission exhibiting a 11.68~d period \citep{1996ApJ...466..384D}. The long term variability of the chromospheric emission was first reported to be chaotic in nature \citep{1995ApJ...438..269B}, before new observations revealed that the magnetic activity has become more regular during the last two decades, following a 2.95~yr chromospheric cycle \citep{2013ApJ...763L..26M}. This cyclic pattern lasted until about 2017, with less regular variations reported after this date \citep{2020A&A...636A..49C}. The surface magnetic field of \epseri\ was first detected through Zeeman broadening at infrared wavelengths \citep{1995ApJ...439..939V} and then in the visible domain \citep{1997A&A...318..429R}. Zeeman-Doppler Imaging (ZDI hereafter) was applied to spectropolarimetric observations of \epseri, with monitoring of the evolution of the large-scale magnetic geometry over timescales as long as several years \citep{2014A&A...569A..79J} and as short as a few months \citep{2017MNRAS.471L..96J}.  

We present here quasi-simultaneous observations of this prototypical young, active solar-type star using spectropolarimetric data collected from the ground with SPIRou and NARVAL, and space-borne photometric data delivered by the TESS spacecraft. We first describe the data sets (Sec. \ref{sec:obs}), then present our determination of fundamental stellar parameters (Sec. \ref{sec:param}), as well as the measurements and modeling of the longitudinal magnetic field (Sec. \ref{sec:bl}), Zeeman broadening (Sec. \ref{sec:broadening}), chromospheric emission (Sec. \ref{sec:sindex}), radial velocities (Sec. \ref{sec:rv}), brightness fluctuations (Sec. \ref{sec:bright}), large-scale magnetic geometry (Sec. \ref{sec:zdi}) and surface differential rotation (Sec. \ref{sec:drot}). We finally summarize, compare and discuss these different measurements (Sec. \ref{sec:discussion}).

\section{Observations}
\label{sec:obs}

\begin{table*}
\caption{Journal of SPIRou and NARVAL observations. We list here the Julian date, rotational phase, exposure time, stellar effective radial velocity, longitudinal magnetic field, and the S-index. RV and $B_{\rm eff}$ measurements with SPIRou are obtained from a night average of LSD profiles. Typical uncertainties for NARVAL radial velocities and S-index measures are 30~\ms\ and 0.002, respectively. The RV precision refers to the relative RV, not the absolute RV. There is likely to be a systematic RV shift (larger than the SPIRou RV error bar) between the two instruments.} 
\centering
\begin{tabular}{c c c c c c c}
\hline
Instrument & HJD & phase & exp. time (sec) & RV (\kms) & $B_{\rm eff}$ (G) & S-index\\
\hline
SPIRou & 2458384.1653  &  0.6991 & $5 \times 4 \times 11.1$ & $16.441 \pm 0.002$ & $ -2.6 \pm 0.8 $ & -- \\
 & 2458385.1645  &  0.7846 & $4 \times 4 \times 33.4$ & $16.432 \pm 0.001$ & $-3.1 \pm 1.0$ & -- \\
 & 2458387.1576 & 0.9553 & $5 \times 4 \times 33.4$ & $16.451 \pm 0.001$ & $+0.6 \pm 0.8$ & -- \\
 & 2458388.1564 & 0.0408 & $1 \times 4 \times 33.4$ & $16.456 \pm 0.002$ & $+2.5 \pm 1.0$ & -- \\
\hline
\hline
NARVAL &  2458380.6620  &  0.3992 & $4 \times 200$  &  16.53  & $ -2.7  \pm  0.2 $ &  0.495  \\
 &  2458381.6358  &  0.4825 & $4 \times 200$  &  16.54  & $ -2.8  \pm  0.3 $ &  0.479  \\
 &  2458384.6252  &  0.7385 & $4 \times 200$  &  16.54  & $ -4.2  \pm  0.2 $ &  0.428  \\
 &  2458386.6208  &  0.9093 & $4 \times 200$  &  16.51  & $ -0.6  \pm  0.3 $ &  0.444  \\
 &  2458387.6261  &  0.9954 & $4 \times 200$  &  16.52  & $ +0.4  \pm  0.2 $ &  0.452  \\
 &  2458388.6513  &  0.0832 & $4 \times 200$  &  16.50  & $ +1.1  \pm  0.2 $ &  0.459  \\
 &  2458389.5919  &  0.1637 & $4 \times 200$  &  16.52  & $ +0.9  \pm  0.2 $ &  0.479  \\
 &  2458395.6595  &  0.6832 & $4 \times 200$  &  16.46  & $ -4.8  \pm  0.3 $ &  0.446  \\
 &  2458396.6073  &  0.7643 & $4 \times 200$  &  16.46  & $ -3.3  \pm  0.2 $ &  0.443  \\
 &  2458397.6046  &  0.8497 & $4 \times 200$  &  16.45  & $ -2.0  \pm  0.2 $ &  0.439  \\
 &  2458400.5678  &  0.1034 & $4 \times 200$  &  16.51  & $ -0.3  \pm  0.5 $ &  0.447  \\
 &  2458404.5693  &  0.4460 & $4 \times 200$  &  16.52  & $ -2.5  \pm  0.4 $ &  0.465  \\
 &  2458414.5792  &  0.3030 & $4 \times 200$  &  16.48  & $ +2.3  \pm  0.2 $ &  0.473  \\
 &  2458415.5853  &  0.3892 & $4 \times 200$  &  16.50  & $ +0.9  \pm  0.3 $ &  0.475  \\
 &  2458416.5284  &  0.4699 & $4 \times 200$  &  16.50  & $ -3.0  \pm  0.2 $ &  0.451  \\
 &  2458417.5236  &  0.5551 & $4 \times 200$  &  16.49  & $ -4.9  \pm  0.2 $ &  0.452  \\
 &  2458430.4804  &  0.6644 & $4 \times 200$  &  16.47  & $ -2.8  \pm  0.3 $ &  0.439  \\
 &  2458436.5098  &  0.1806 & $4 \times 200$  &  16.53  & $ +3.0  \pm  0.2 $ &  0.444  \\
 &  2458437.4856  &  0.2642 & $4 \times 200$  &  16.52  & $ +2.3  \pm  0.2 $ &  0.436  \\
 &  2458439.4952  &  0.4362 & $4 \times 200$  &  16.51  & $ -2.1  \pm  0.2 $ &  0.433  \\

\hline
\end{tabular}
\label{tab:journal}
\end{table*}

We report on three time-series of observations of \epseri\ taken with three different instruments in 2018, over a period of about two months spread from late September to late November. Spectropolarimetric data were obtained in the visible and near-infrared (nIR) domain, using NARVAL and SPIRou, respectively. Photometric monitoring was performed onboard the TESS spacecraft.  

\subsection{SPIRou near-infrared spectra}

SPIRou is a cryogenic, near-infrared, high-resolution \'echelle spectropolarimeter and velocimeter recently installed at CFHT \citep{2020MNRAS.tmp.2502D}. Each spectrum covers the Y, J, H, and K bands (nominal spectral range from 0.98 to 2.35 $\mu$m), at a spectral resolution of 70,000, with a radial velocity precision of between 1 and $2$ \ms\ RMS. Similar to its optical predecessors ESPaDOnS and NARVAL, Stokes $V$ sequences produced by SPIRou consist of four sub-exposures collected with different rotation angles of the half-wave Fresnel rhombs in the Cassegrain mounted polarimeter. This procedure ensures that the two polarimetric states exchange their position on the H4RG detector,  so that spurious polarimetric signatures can be removed at first order \citep{1993A&A...278..231S,donati97}. 

The data reduction was performed with an adapted version of the \textsc{LibreEsprit} pipeline \citep{donati97}. The spectral domain of SPIRou being affected by a large number of telluric lines, their subtraction was performed with a Principal Component Analysis approach inspired by \cite{2014SPIE.9149E..05A}, using a large number of observations of hot stars (with very few photospheric lines in the near  infrared) as a learning data set featuring a variety of configurations of the telluric spectrum.

Observations of \epseri\ were obtained as part of the science verification of the instrument. One to five Stokes $V$ spectra were obtained daily between 21 Sep and 25 Sep, with one missing night over this period (23 Sep). The integration time, first set to 11.1~sec per sub-exposure on Sep 21, was then increased to 33.4~sec per sub-exposure. The peak signal-to-noise ratio (S/N hereafter) in Stokes $I$ spectra ranges from about 800 on 21 Sep to 1,300 during the following nights (this peak value is reached at a wavelength of around 2.1 $\mu$m). The polarimetric sequence was obtained only once on 25 Sep, and was repeated four times on 22 Sep and five times on 21 Sep and 24 Sep (for a total of 15 spectra over the four nights). 

\subsection{NARVAL visible spectra}

NARVAL is a high-resolution \'echelle spectropolarimeter installed at T\'elescope Bernard Lyot (Pic du Midi Observatory, France, \citealt{auriere03}). Its spectral resolution in polarimetric mode is close to 65,000, and it is able to cover, in a single exposure, a spectral domain ranging from 370 nm to 1,000 nm (except for small wavelength gaps between the redmost spectral orders). We have collected here circularly polarized sequences only (Stokes parameters I and V), because the amplitude of Zeeman signatures is maximal in this polarization state. All spectra were reduced by the automated \textsc{LibreEsprit} pipeline \citep{donati97}.

The NARVAL data set is constituted of 20 visits to the target, obtained as part of the Bcool Large Program \citep{2014MNRAS.444.3517M}. This observing sequence of observations started on Sep. 18, and we accumulated new data until Nov. 16. Four NARVAL spectra were obtained during the SPIRou observing window, and 6 spectra are collected during the TESS monitoring (see Sec. \ref{sec:TESS}).

The choice to adopt integration times of $4\times200$~sec in polarimetric sequences results in  peak \sn\ (reached in order \#31, at about 730 nm) mostly comprised between 1,200 and 2,000. Only two spectra, taken on Oct. 08 and 12, have their peak \sn\ slightly below 1,000. The collected temporal sequence suffers from a few gaps, the most extended one spanning 13 nights between Oct 25 and Nov 07 (immediately followed by a new 5 night gap).

All NARVAL data used here are publicly available on the PolarBase archive \citep{2014PASP..126..469P}.

\subsection{TESS photometry}
\label{sec:TESS}

The Transiting Exoplanet Survey Satellite (TESS, \citealt{ricker15}) performed observations of \epseri\ as part of the Sector 4 sequence, lasting from Oct 20 to Nov 19 (HJD 2458410.9 to 2458436.8). The public, reduced light curve was retrieved from the MAST archive. 

\begin{figure*} 
\includegraphics[width=18cm]{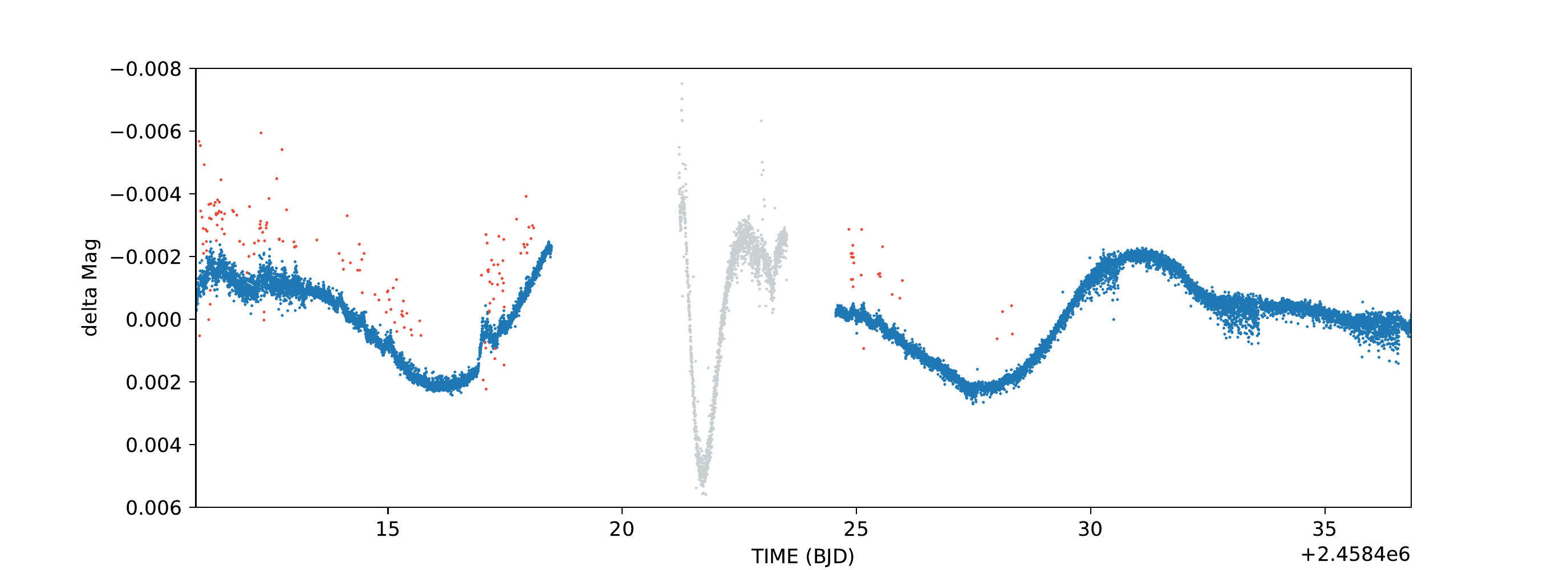} 
\caption{Normalized TESS light curve of \epseri, as a function of the barycentric Julian date. Red points show the observations affected by a non-zero quality flag. Grey points were impacted by measurement instabilities that were also recorded on other targets observed by the same detector. Blue points are the ones retained for our study.}
\label{fig:TESS}
\end{figure*}

Each data point in the light curve is delivered with a 12-bit quality flag, with each bit indicating an abnormal data condition if positive. All the flagged data points were discarded from our study. We also rejected data obtained between Julian dates 2458420 and 2458424 because (a) this fraction of the light curve exhibited a periodicity and amplitude inconsistent with the rest of the time series, and (b) other light curves of stars on the same detector (e.g. $\delta$~Eri, 7~Eri, HD~19349) also show a similarly abnormal behaviour at the same dates, which lead us to conclude that TESS was undergoing sub-nominal conditions while taking these measurements. The resulting time series is a constituted of 13989 points (Fig \ref{fig:TESS}).

\section{Fundamental parameters}
\label{sec:param}

\begin{figure*} 
\includegraphics[width=13cm,angle=90]{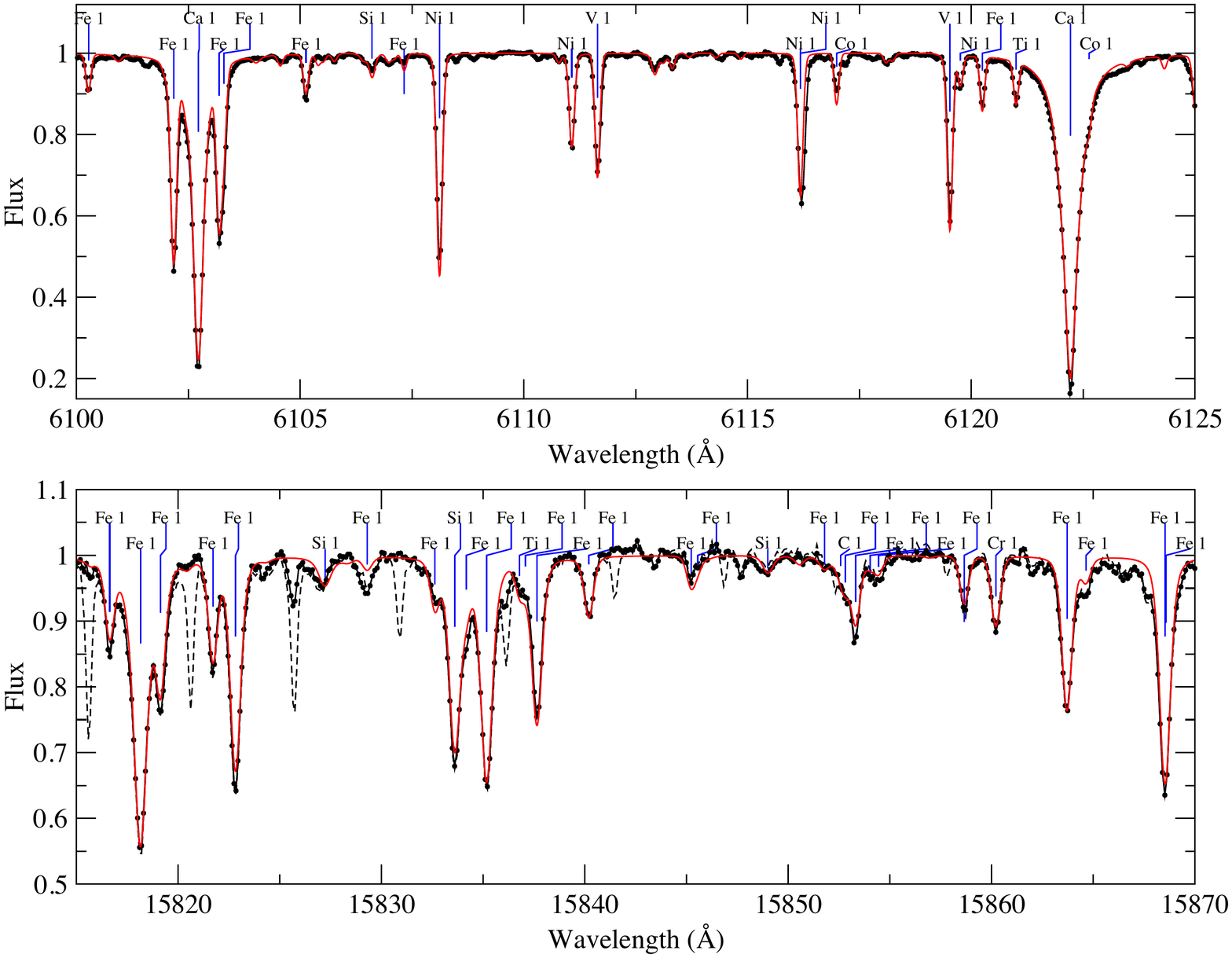} 
\caption{{\sc Zeeman} adjustment (red) of spectral lines in the SPIRou wavelength domain after removal of telluric lines (black line). The dashed line shows the SPIRou spectrum prior to the subtraction of tellurics. Vertical blue ticks show the wavelength position of spectral lines included in the {\sc Zeeman} analysis.}
\label{fig:specfit}
\end{figure*}

The physical parameters of $\epsilon$ Eri are generally well known.  There have been numerous spectroscopic studies, some reaching very good precision \citep[e.g.][]{2005ApJS..159..141V}.  Furthermore, there are two interferometric diameter estimates (\citealt{diFolco2007}, and  \citealt{2012ApJ...744..138B}), providing independent information.  The star was included in the meta-analysis of \citet{heiter2015} of Gaia FGK benchmark stars.  In this context, an additional spectroscopic analysis is not urgent, however with both visible and IR spectra we can perform a detailed comparison between parameters derived in widely separated spectral regions using contemporaneous observations.  

Our analysis proceeded by directly fitting synthetic spectra to the observations by $\chi^2$ minimization.  The stellar parameters fit were effective temperature (\teff), surface gravity (\logg), macroturbulence (\vmac), microturbulence (\vmic), metallicity and radial velocity.  Line broadening from macroturbulence and \vsin\ is ambiguous for such a slow rotator, thus we used the interferometric radius of \citet[][$0.74 \pm 0.01 R_\odot$]{diFolco2007} and the equatorial rotation period and inclination of \citet[][ \peq~$= 10.33$ d, $i=46^\circ$]{2014A&A...569A..79J} to infer a \vsin\ of 2.59 \kms.
We calculated synthetic spectra using the {\sc Zeeman} spectrum synthesis code \citep{Landstreet1988, Wade2001}.  This calculates atomic line spectra in LTE, using plane-parallel model atmospheres.  Input model atmospheres from the MARCS grid of \citet{Gustafsson2008} were used.  Input atomic data from VALD \citep{Piskunov1995-VALD,Ryabchikova1997-VALD,Kupka1999-VALD,Kupka2000-VALD,Ryabchikova2015-VALD} were used. The analysis proceeded by fitting several spectroscopic windows independently, then taking the average as the best value and the standard deviation as the uncertainty on that value. This may overestimate the uncertainty, since the standard deviation may be driven by the worst spectroscopic window.  However, we prefer it to simply using the covariance matrix, since this more completely accounts for possible systematic errors in the model and atomic data.

To achieve a high degree of precision in spectrum modelling across many lines, empirical corrections to atomic line oscillator strengths are commonly needed.  We find this is particularly true in the IR, possibly due to this less commonly used data being less accurate, or possibly due to the lower density of lines per nm causing errors to average out less effectively.
We derive empirical oscillator strength corrections by modelling the solar spectrum.  We used observations of sunlight reflected from the moon with NARVAL and SPIRou.  A synthetic spectrum was calculated for solar parameters, then discrepant lines were identified by hand, and oscillator strengths for those lines were iteratively fit, further discussed in Sec.\ \ref{sec:param_SPIRou}.  

\subsection{NARVAL visible analysis}

For the analysis in the visible we used the NARVAL spectrum from the night of Sep.\ 25, 2018.  The Stokes $I$ spectrum did not vary noticeably and the \sn\ was high enough that coadding spectra was unnecessary.  The fitting procedure closely followed \citet{2016MNRAS.457..580F} and \citet{2018MNRAS.474.4956F}.  We used six spectral windows each roughly 100 \AA\ long: 6025--6100, 6100--6200, 6200--6275, 6314--6402, 6402--6500, and 6590--6700 \AA.  Regions contaminated with telluric lines, as well as a few other features not present in our line list, were avoided. The Li line at 6707.8 \AA\ was not unambiguously detectable in the observation, so a reliable Li abundance could not be derived.
We used a fixed \vsin\ of 2.59 \kms\ and fit macroturbulence. However, if we were to assume no macroturbulence and fit for \vsin\ we would get $2.87 \pm 0.26$ \kms, which is consistent with the value based on the rotation period and radius plus a small amount of turbulent broadening.
The resulting best fit stellar parameters averaged over the windows are presented in Tab. \ref{table-phys-params}.  Our results are generally in good agreement with the results of \citet{2005ApJS..159..141V}, and \citet{heiter2015}. 

\begin{table*}
\centering
\caption{Physical parameters for $\epsilon$ Eri derived from visible and IR spectra.
References for literature values: $^1$ \citet{heiter2015}, $^2$ \citet{2005ApJS..159..141V}, $^3$ \citet{Luck2005}, $^4$ \cite{Jofre2014}. }
\begin{tabular}{lccc}
\hline
   & visible & IR & literature \\
\hline
\teff\ (K)    & $ 5010 \pm 64  $ & $ 4991 \pm 59  $ & $ 5076 \pm 30  ^1$ \\
\logg\  (cgs) & $ 4.53 \pm 0.08$ & $ 4.48 \pm 0.14$ & $ 4.61 \pm 0.03^1$ \\
\vsin\ (\kms) & $ 2.59 $         & $ 2.59$          & $ 2.4  \pm 0.5 ^2$ \\
\vmic\ (\kms) & $ 0.99 \pm 0.13$ & $ 1.18 \pm 0.22$ & $0.7^3$, $1.14\pm0.05^4$ \\
\vmac\ (\kms) & $ 0.93 \pm 0.38$ & $ 3.03 \pm 0.85$ & \\
metallicity   & $-0.08 \pm 0.04$ & $-0.09 \pm 0.04$ & $-0.09 \pm 0.06^{4,1}$ \\
\hline
\end{tabular} 
\label{table-phys-params} 
\end{table*}

\subsection{SPIRou infrared analysis}
\label{sec:param_SPIRou}

For the analysis in the IR we used the SPIRou spectra from 21 Sep.\ and co-added them to produce one high \sn\ spectrum for the night, using the telluric correction provided by our upgraded \textsc{LibreEsprit} pipeline.  We found no clear variability between nights in the absorption lines, and results of this analysis from different nights were consistent within the noise.  Typically there are fewer spectral lines per \AA\ in the IR than in the visible for a K star, so we used larger spectral windows.  Since regions of the spectrum that are dominated by telluric lines are very difficult to fully correct, such regions were avoided entirely.  This lead to using large windows of: 10500--10920, 11760--12600, 15110--15697, 15815--16390, 16439--17140, and 21017--22850 \AA.  This spans most of the usable spectral range of SPIRou, in 6 independent windows.  Within these windows there were several gaps, avoiding stronger telluric features (most notably from 15985 to 16145 \AA), a few broad features not present in our model spectra, and stronger molecular lines.  
Comparing the telluric corrected and uncorrected spectra was very helpful for identifying regions where the telluric correction may not have been sufficiently accurate, and such regions were avoided.

A major limitation of {\sc Zeeman} is that it does not currently compute molecular lines in stellar spectra.  In the visible for K stars molecular lines can easily be identified and avoided, particularly since VALD version 3 contains extensive molecular line lists.  However in some regions of the SPIRou domain this becomes more difficult, even for a star of $\sim$5000 K.  Care was taken to identify and avoid molecular features, however a few very weak lines of CN and OH could not be avoided in the 11760--12600, 15815--16390, and 16439--17140  \AA\ regions, and a larger number of very weak CN lines were unavoidable in the 15110--15697 \AA\ region. (A number of CN lines are also present in the 21017--22850 \AA\ region but they can be more practically avoided due to the large spacing between lines).  Since these very weak lines are at or near the noise level, we do not expect them to have a large impact on the results, but they may contribute to our uncertainties by increasing the standard deviation of results from different windows.  The 15110--15697 \AA\ region may be viewed as the least reliable due to the larger number of contaminating molecular lines.  

The atomic line data in the SPIRou domain appears to be less reliable than in the visible, or at least obtaining accurate results is more reliant on corrections to the input atomic data.  There is a good agreement on the presence of lines in VALD and in the observation, with a few exceptions, and the wavelengths generally appear to be correct.  However line strengths are in many cases wrong, and in some stronger lines the width of the Lorentzian component of the line also appears to be incorrect.  
To correct for this, we used a spectrum of sunlight reflected off the Moon obtained with SPIRou.  We computed synthetic spectra with solar parameters (\teff\ $=5777$ K, \logg\ $=4.4$ (with $g$ in cm s$^{-1}$), \vsin\ $=2$ \kms, \vmic\ $=0.9$ \kms, and macroturbulence of 3 \kms), and identified discrepant lines visually.  We then fit the oscillator strength ($\log gf$) of the lines by $\chi^2$ minimization.  For some lines it was also necessary to fit the van der Waals broadening coefficient (van der Waals broadening is typically the dominant Lorentzian broadening source in K stars), and it was included in a simultaneous fit with $\log gf$.  These empirical corrections were essential for deriving accurate stellar parameters and achieving reasonable fits to the infrared observations.  Corrections for 480 lines were used in our final results, as detailed in Appendix \ref{sec:linecorr}.  To verify the reliability of these empirical corrections, we fit the solar spectrum using the corrections to derive \teff, \logg, \vmic, macroturbulence and metallicity, and ensured we obtained results consistent with the solar values.  

The best fit parameters for individual windows in the SPIRou spectrum are presented in Tab. \ref{table-params-windows}, and the final average and standard deviation are in Tab. \ref{table-phys-params}.  We find a good agreement in \teff, \logg, and metallicity across all six spectral windows in the infrared, and a very good agreement between the average results for the visible and infrared spectra. For comparison we include well established literature values in Tab. \ref{table-phys-params}, and find good agreement within our uncertainty.  For microturbulence, there are two windows where we find anomalously low values, this may be due to the influence of weak unidentified blends or incompletely removed telluric lines.  Rejecting the most extreme microturbulence value as likely wrong, we obtain an average that is in good agreement with the optical and literature values, albeit with a larger uncertainty. The larger macroturbulence value found in the infrared is discussed in Appendix \ref{sec:macro}. 

\section{Longitudinal field measurements}
\label{sec:bl}

The Least-Square Deconvolution method (LSD hereafter, \citealt{donati97,kochukhov10}), was applied to every NARVAL and SPIRou reduced spectrum to get the S/N boost necessary to the detection of polarized Zeeman signatures and to precise radial velocity measurements. We used a visible and infrared list of photospheric spectral lines computed for effective temperature and surface gravity closely matching the stellar parameters derived in Sec. \ref{sec:param} (and ignoring portions of the spectra blended with chromospheric lines or heavily affected by telluric absorption). The adopted lists feature close to 8,000 lines in the SPIRou domain, and 11,000 lines in the visible domain. The outcome of this procedure is a single, pseudo line profile with a typical \sn\ in Stokes $V$ as large as 18,000 for SPIRou observations, versus about 50,000 for NARVAL data.   

The projection on the line of sight of the disc-integrated magnetic field ($B_{\rm eff}$) can be estimated through the first moment of Stokes $V$ LSD profiles, following \cite{1979A&A....74....1R,donati97}:  

\begin{equation}
B_{\rm eff} = -2.14\times10^{-11}\frac{\int v V(v)dv}{\lambda_{0} g_{\rm eff} c \int (1-I(v))dv}~G.
\label{bl_measure}
\end{equation}

\noindent where $v$ is the radial velocity, $\lambda_{0}$ is the central wavelength of LSD pseudo-profiles, and  $g_{\rm eff}$ its effective Land\'e factor. 

\begin{figure} 
\includegraphics[width=\linewidth]{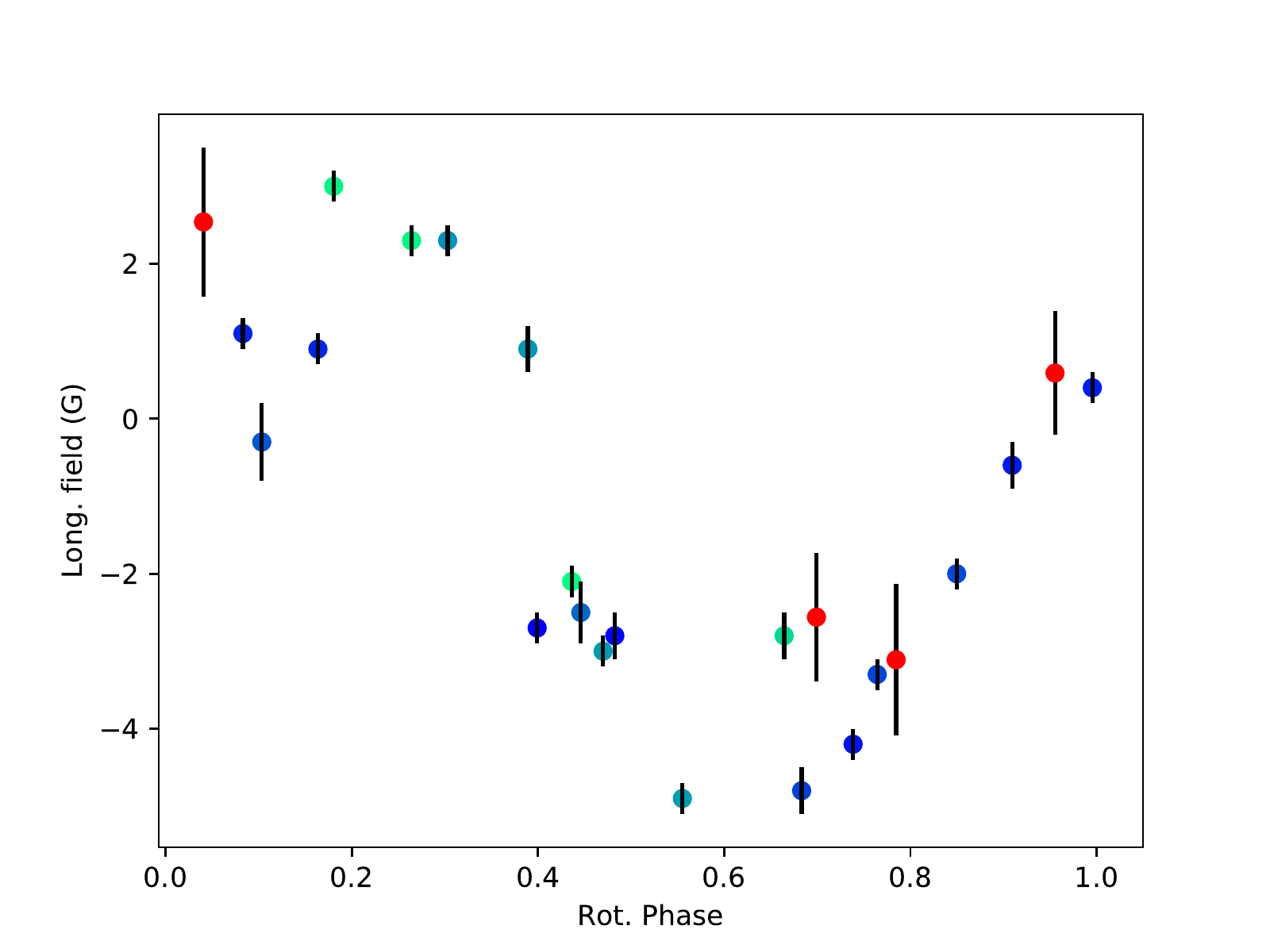} 
\caption{Longitudinal magnetic field measurements, as a function of the rotational phase. Red symbols show SPIRou estimates. Blue/green symbols correspond to NARVAL measurements, with bluer (resp. greener) symbols showing older (resp. more recent) observations.}
\label{fig:bl}
\end{figure}

The series of $B_{\rm eff}$ measurements is plotted  in Fig. \ref{fig:bl}, as a function of the rotational phase. The phases were computed  using a rotation period equal to 11.68~d \citep{1996ApJ...466..384D}, and the null phase is set at HJD=2458399.36, which corresponds to the mean date of the NARVAL observations. We note that the four SPIRou measurements are consistent with NARVAL estimates obtained at close-by phases, although the shorter exposure times, smaller number of lines used to compute the SPIRou LSD pseudo-profiles, and lower line depth in the nIR lead to larger error bars. 

The rotational modulation of $B_{\rm eff}$ is obvious, with a smooth transition from values close to -5~G around phase 0.6 to +3~G at phase 0.2. Some scatter is also observed, showing that the temporal variability is not entirely controlled by stellar rotation. We note that a fair part of the scatter is due to the most recent NARVAL observations, with differences as large as 2~G between the oldest and most recent observations, with a global trend to get larger field estimates (negative fields becoming weaker and positive fields becoming stronger) in more recent data.

To model both the rotational modulation and its temporal evolution, we fit the series of $B_{\rm eff}$ measurements by means of Gaussian Process Regression (GPR, \citealt{2014MNRAS.443.2517H,2017MNRAS.467.1342Y}). We use here a pseudo-periodic co-variance function, defined for times $t$ and $t'$ and for the four hyperparameters $\theta_1$ to $\theta_4$ by the following equation:

\begin{equation}
    K(t,t') = \theta_1^2\exp\left[-\frac{(t - t')^2}{\theta_3^2} - \frac{\sin^2\left(\frac{\pi(t - t')}{\theta_2}\right)}{\theta^2_4}\right]
\end{equation}
\label{eq:gpr}

\noindent where $\theta_1$ is the amplitude (in Gauss) of the GP, $\theta_2$ the cycle length ($i.e.$ the rotation period, in days), $\theta_3$ the decay timescale (in  days, describing the lifetime of magnetic spots contributing  to the large-scale magnetic geometry), and $\theta_4$ a smoothing  parameter in the [0,1] interval. 

\begin{figure} 
\includegraphics[width=9cm,angle=-90]{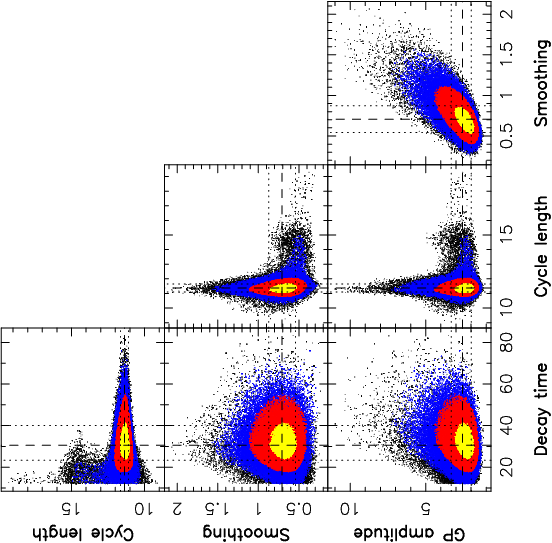} 
\caption{Outcome of the MCMC run for the four GP hyperparameters of the $B_{\rm eff}$ model. The decay time and cycle length are expressed in days, the GP amplitude in Gauss, and the smoothing parameter is dimensionless. Dashed lines show the best parameters, while dotted lines indicate the error bars.}
\label{fig:MCMC}
\end{figure}


\begin{table}
\centering
\caption{Priors used for the MCMC exploration of the hyperparameter space.}
\begin{tabular}{lll}
\hline
Hyperparameter & Prior type & adopted value   \\
\hline
$\theta_1$ (amplitude) & Modified Jeffreys & MJ$(0.01,\, 700)$\\
$\theta_2$ (rotation period) &  Gaussian & $\mathcal{N}(11.5,\,0.1)$\\
$\theta_3$ (decay time) &  Jeffreys & J$(12,\, 700)$ \\
$\theta_4$ (smoothing) &  Uniform & $\mathcal{U}[0.001,\,1]$ \\
\hline
\end{tabular} 
\label{table-MCMC} 
\end{table}

The hyperparameter domain is explored through the Markov Chain Monte Carlo (MCMC) method. The result of the MCMC simulation is illustrated in Fig. \ref{fig:MCMC}. The resulting cycle length is equal to $11.4 \pm 0.3$~d, with an amplitude of $2.6^{+0.7}_{-0.6}$~G, a decay time equal to $31^{+9}_{-7}$~d, and a smoothing parameter of $0.7 \pm 0.2$. 

\section{Zeeman broadening}
\label{sec:broadening}

\begin{figure*} 
\centering
\includegraphics[width=13cm,angle=90]{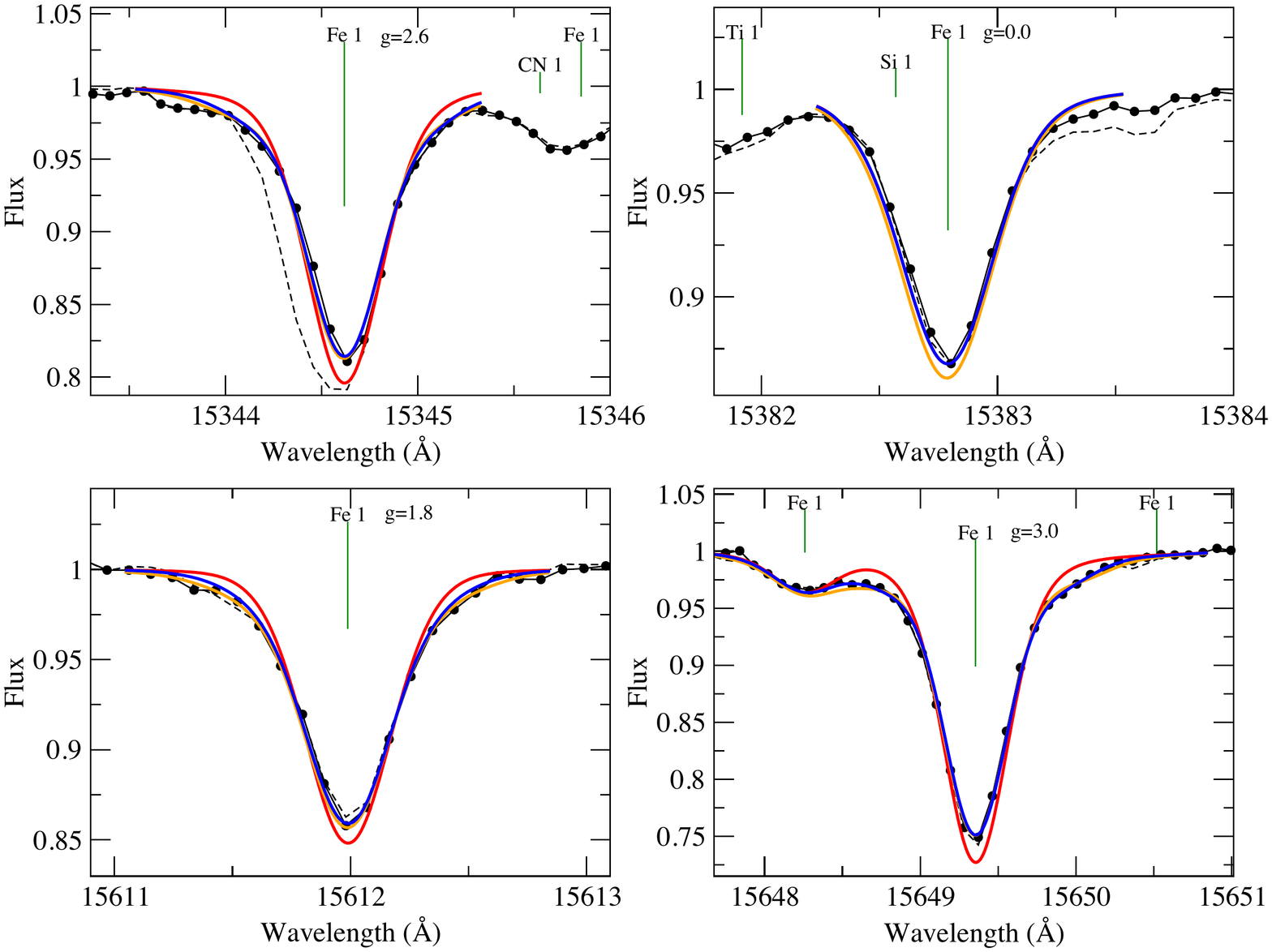} \caption{Infrared spectral lines of \epseri\ observed with SPIRou (black points) and modelled with {\sc Zeeman}. The wavelength of lines is shown as green vertical ticks, and the effective Land\'e $g$ factors are indicated to the right of each modelled line name. The best two-component magnetic model (blue) provides an acceptable fit to all lines, while a model with no magnetic field but otherwise the same parameters (red) only fits the magnetically insensitive line (upper right panel). A multi-component model (orange) provides a comparable fit to the two-component model. } The observation before telluric correction (dashed line) is shown for reference.
\label{fig:zeeman_broad}
\end{figure*}

The strength of the stellar magnetic field can be estimated from the impact of Zeeman splitting in Stokes $I$.  Stokes $V$ observations are sensitive to the sign of the line of sight component of the magnetic field, thus they are relatively sensitive to the geometry of the magnetic field, but nearby regions of opposite sign can cancel out, effectively becoming undetectable.  Stokes $I$ is relatively insensitive to geometry, being sensitive to the strength but not the orientation of the magnetic field\footnote{Zeeman broadening in Stokes $I$ also typically lacks Doppler resolution across the surface of the star, since it is only reliably detectable when it is comparable to, or larger than, the rotational broadening.  This is another important difference with Stokes $V$ measurements, which can still reliably detect Zeeman splitting when it is much less than rotational or other line broadening processes.}.  Thus for a magnetic field distribution similar to that of the Sun, Stokes $I$ provides a measure of small intense magnetic field regions that cover a small fraction of the stellar surface and largely cancel out in Stokes $V$.  Since the Zeeman effect scales as $\lambda^2$ while most other broadening processes scale with $\lambda$, lines further into the infrared provide a more sensitive diagnostic.  The approach we take here is to directly fit a small number of carefully selected lines with synthetic spectra computed by {\sc Zeeman}.  

This method relies on a selection of good lines with a range of Land\'e factors, and the wide spectral domain of SPIRou offers many possibilities.  We looked for lines with particularly large or small effective Land\'e factors ($g_{\rm eff}$), minimal blending, preferred wavelengths longer than 15000 \AA, and avoided anything blended with strong telluric lines in case of imperfections in the telluric correction.  We focused on lines that are strong enough to limit the impact of noise, but not so strong that they have large Lorentzian wings.  The \sn\ of the observation begins to decrease beyond 2 $\mu$m, so lines in the K band are less optimal.  

We adopt four Fe {\sc i} lines for this analysis, at 15343.79 ($g_{\rm eff}=2.63$), 15381.96 ($g_{\rm eff}=0.01$), 15611.14 ($g_{\rm eff}=1.83$), and 15648.51 ($g_{\rm eff}=2.98$) \AA.  The Fe {\sc i} 15381.96 \AA\ line has an effective Land\'e factor near zero, and provides a magnetically insensitive diagnostic for turbulent and rotational broadening.  The Fe {\sc i} 15648.51 and 15343.79 \AA\ lines have exceptionally large effective Land\'e factors, while the 15611.14 line is also quite sensitive.  The Fe {\sc i} 15648.51 \AA\ line has been used by several other authors \citep[e.g.][]{1995ApJ...439..939V, lavail2017}.  

Some alternate lines that were considered but not used include the Fe {\sc i} 15534.3 \AA\ line \citep{1995ApJ...439..939V}, which has a large effective Land\'e (1.95) factor but may have weak CN blends in the red wing, so we excluded it.  Similarly, the very low Land\'e factor Fe {\sc i} 15560.8 \AA\ line \citep{1995ApJ...439..939V} is weak and appears to have a weak OH or CN blend in the blue wing making it unsuitable.  
The Fe {\sc i} 15621.65 and 15662.0 \AA\ lines \citep{lavail2017} are strong with large Loretzian wings in \epseri, which makes distinguishing pressure and Zeeman broadening more difficult, thus they were not used.  
The Ti {\sc i} 22211.2, 22232.9, 22274.0, 22310.6 \AA\ lines \citep{JohnsKrull2004} have large effective Land\'e factors (up to 2.5 for 22310.6 \AA), but are relatively weak and in a lower SNR portion of the \epseri\  spectrum, and in some observations can be blended with strong telluric lines. Thus they would be good choices for cooler stars, but are not optimal for \epseri.

\begin{table}
\centering
\caption{Atomic line data for the Fe {\sc i} lines used in the Zeeman broadening analysis, including the effective Land\'e $g$ factor and van der Waals damping parameter. }
\begin{tabular}{cccc}
\hline
$\lambda$ (\AA) & $\log gf$ & $g_{\rm eff}$ & van der Waals \\
\hline
15343.788 & $-0.67 \pm 0.04$  & 2.63 & -7.52 \\
15381.960 & $-0.69 \pm 0.04$  & 0.01 & -7.43 \\
15611.145 & $-3.30 \pm 0.07$ & 1.83 & -7.79 \\
15647.413 & $-1.08 \pm 0.11$  & 1.00 & -7.29 \\
15648.510 & -0.63 & 2.98 & -7.49 \\
\hline
\end{tabular} 
\label{table-line-data-zbroad} 
\end{table}


\begin{table*}
\centering
\caption{Best fit Zeeman broadening results for \epseri.  Results are presented for the two-component model, and for a multi-component model that provides limits on the presence of stronger magnetic fields.  Parameters without uncertainties were fixed based on a fit to the visible spectrum.  The final column contains averages over the four nights, with the standard deviation as the uncertainty.}
\begin{tabular}{lccccc}
\hline
                   & \multicolumn{5}{c}{two-component model} \\
parameter          & 21 Sept.          & 22 Sept.          & 24 Sept.          & 25 Sept.          & average           \\
\hline
\teff\ (K)         & \multicolumn{5}{c}{5010}  \\
\logg\  (cgs)      & \multicolumn{5}{c}{4.53}  \\
\vsin\ (\kms)      & \multicolumn{5}{c}{2.59}  \\
\vmic\ (\kms)      & \multicolumn{5}{c}{0.99}  \\
\vmac\ (\kms)      & $4.93 \pm 0.24$   & $3.03 \pm 0.24$   & $3.34 \pm 0.18$   & $3.19 \pm 0.22$   & $3.26 \pm 0.17$   \\
metallicity        & $0.038 \pm 0.007$ & $0.024 \pm 0.010$ & $0.015 \pm 0.008$ & $0.012 \pm 0.009$ & $0.022 \pm 0.010$ \\
$B_{\rm mono}$ (G)   & $1898  \pm 129  $ & $1812  \pm 136  $ & $1846  \pm 133  $ & $1783  \pm 135  $ & $1835  \pm 43  $  \\
filling factor     & $0.125 \pm 0.017$ & $0.159 \pm 0.023$ & $0.126 \pm 0.018$ & $0.141 \pm 0.021$ & $0.138 \pm 0.014$ \\
\hline
                   & \multicolumn{5}{c}{multi-component model} \\
parameter          & 21 Sept.          & 22 Sept.          & 24 Sept.          & 25 Sept.          & average           \\
\hline
\teff\ (K)         & \multicolumn{5}{c}{5010}  \\
\logg\  (cgs)      & \multicolumn{5}{c}{4.53}  \\
\vsin\ (\kms)      & \multicolumn{5}{c}{2.59}  \\
\vmic\ (\kms)      & \multicolumn{5}{c}{0.99}  \\
\vmac\ (\kms)      & $5.02 \pm 0.21$   & $3.22 \pm 0.19$   & $3.44 \pm 0.19$   & $3.40 \pm 0.18$   & $3.40 \pm 0.12$   \\
metallicity        & $0.048 \pm 0.012$ & $0.029 \pm 0.016$ & $-0.008\pm 0.015$ & $0.005 \pm 0.014$ & $0.019 \pm 0.022$ \\
$f_{\rm 0 kG}$       & $0.831 \pm 0.043$ & $0.840 \pm 0.057$ & $0.850 \pm 0.054$ & $0.859 \pm 0.051$ & $0.845 \pm 0.011$ \\
$f_{\rm 2 kG}$       & $0.118 \pm 0.017$ & $0.131 \pm 0.023$ & $0.140 \pm 0.022$ & $0.129 \pm 0.020$ & $0.130 \pm 0.008$ \\
$f_{\rm 4 kG}$       & $0.000 \pm 0.022$ & $0.028 \pm 0.028$ & $0.009 \pm 0.026$ & $0.010 \pm 0.025$ & $0.012 \pm 0.010$ \\
$f_{\rm 6 kG}$       & $0.050 \pm 0.033$ & $0.000 \pm 0.044$ & $0.000 \pm 0.042$ & $0.000 \pm 0.040$ & $0.013 \pm 0.021$ \\
\hline
\end{tabular} 
\label{table-zbroad-fit} 
\end{table*}

The oscillator strengths of the four lines we focused on for Zeeman broadening, as well as a weak blending Fe {\sc i} line at 15647.41 \AA, were corrected by fitting synthetic spectra to a solar spectrum, as was done in Sec. \ref{sec:param_SPIRou}.  For the Fe {\sc i} 15648.51 line, we adopt the oscillator strength of \citet{1995ApJ...439..939V}, since it provided an adequate fit for \epseri\ and was consistent with our best fit solar value.  The adopted oscillator strengths, as well as the effective Land\'e factors and van der Waals damping coefficients are provided in Tab. \ref{table-line-data-zbroad}.

The model spectra assume a radial magnetic field uniformly distributed over some fraction of the stellar surface, thus the spectra are effectively the combination of a uniform radial magnetic model and a non-magnetic model.  The fraction of the surface containing magnetic field is parameterized as a filling factor.  This is clearly much simpler than the real magnetic field of a star, but Zeeman broadening in Stokes $I$ is relatively insensitive to the magnetic geometry, so a more realistic geometry would add more free parameters but not improve the fit to the observations.  We use the \teff, \logg, \vsin, and microturbulence derived from the visible range in Sec.\ \ref{sec:param}, however, we allow metallicity and macroturbulence to be free parameters.  This allows the model to fit line strength through metallicity, and line width through macroturbulence, so that assumptions in these parameters do not unduly influence the derived magnetic field strength.  

We consider two approaches to fitting Zeeman broadening parameters. First we consider a model with free parameters for a characteristic magnetic field strength ($B_{\rm mono}$) covering a filling factor ($f$), and the remaining surface is non-magnetic (the `two-component' model).  This is consistent with the approach of, e.g., \citet{Marcy1984}, \citet{Gray1984},  and \citet{1995ApJ...439..939V}, and is reasonable for \epseri\ since Zeeman broadening does not dominate over other broadening processes. Later, we consider a second approach using a model with a grid of fixed magnetic field strengths, and filling factors for those strengths as free parameters (the `multi-component' model, e.g. \citealt{JohnsKrull1999,shulyak19}).  Since the magnetic field of \epseri\ is not extremely strong, we cannot derive an extended distribution of filling factors, but we can place limits on the presence of multi-kilogauss magnetic fields. 

For the coadded spectrum from 21 Sept., the best fit two-component model we find has magnetic field $B_{\rm mono} = 1898 \pm 129$ G with a filling factor of $0.125 \pm 0.017$.  The full set of parameters are presented in Tab. \ref{table-zbroad-fit}, and the fits are presented in Fig.\ \ref{fig:zeeman_broad}, contrasted with a non-magnetic model that fits the Land\'e factor 0.01 line.  The uncertainties are based on the diagonal of the covariance matrix, scaled by the square root of the reduced $\chi^2$, thus they account for noise but not systematic uncertainties in input parameters. 
We repeat this analysis for the nightly coadded spectra for all four nights, and find an average $B_{\rm mono}$ of 1835 G and a standard deviation of 43 G, the average filling factor is 0.138 with a standard deviation of 0.014.  The small standard deviation in $B_{\rm mono}$ suggests we may have overestimated the random uncertainties, although it may simply be due to small number statistics.  We find no statistically significant variation in $B_{\rm mono}$ or $f$ over the four coadded spectra.  

To estimate the impact of uncertainties in other parameters, we try varying the input parameters by a reasonable uncertainty, rerun the fit, and check the impact on the magnetic field and filling factor.  Varying \teff\ by $\pm80$ K changes $B_{\rm mono}$ by $\pm30$ G and the filling factor by  $\pm0.005$, and \logg\ has a smaller impact.  Varying \vsin\ by $\pm0.3$ \kms\ changes $B_{\rm mono}$ by $\pm8$ G and the filling factor by $\pm0.001$, since the free macroturbulence parameter offsets this change.  Changing the microturbulence by $\pm0.15$ \kms\ changes $B_{\rm mono}$ by $\pm10$ G and the filling factor by $\pm0.007$, but has a stronger impact on the inferred metallicity.  If we assume an error in the line oscillator strengths of $\pm0.1$ dex in $\log gf$, the impact is between 50 and 150 G in $B_{\rm mono}$, and 0.02 to 0.03 in filling factor.  The formal uncertainties from fitting the oscillator strengths are less than 0.1 (except for the very weak 15647.413 \AA\ line), but this may still be the dominant source of systematic error.

We then considered a multi-component model, largely to constrain the possibility of very strong magnetic fields covering small areas.  For the multi-component model, we chose a grid of magnetic field strengths of 0, 2, 4 and 6 kG\footnote{For the practical purpose of implementing this within a $\chi^2$ minimization routine, the filling factors for the non-zero field regions are treated as the free parameters and the filling factor for the 0 G region is calculated as $1-\sum_{i} f_{i}$}.  Experiments using a grid spaced only 1 kG apart suggests such small bins are not fully resolved, leading to strong covariances between filling factors and potentially a poorly constrained model.  Fitting the coadded 21 Sept.\ spectrum (see Tab. \ref{table-zbroad-fit}) produces a significant filling factor in the 2 kG bin ($f_{\rm 2 kG} = 0.118 \pm 0.017$), while the filling factors for the 4 kG bin ($f_{\rm 4 kG} = 0.001 \pm 0.022$) and 6 kG bin ($f_{\rm 6 kG} = 0.050 \pm 0.033$) are consistent with zero, and most of the surface is in the 0 kG bin ($f_{\rm 0 kG} = 0.831 \pm 0.043$).  The coadded 22 Sept.\ spectrum produces results consistent within $1\sigma$ with $f_{\rm 0 kG} = 0.840 \pm 0.057$, $f_{\rm 2 kG} = 0.131 \pm 0.023$, $f_{\rm 4 kG} = 0.028 \pm 0.028$, and $f_{\rm 6 kG} = 0.000 \pm 0.044$, and there is no statistically significant variability on the Sept.\ 24 or Sept.\ 25.  From this we find no evidence for a magnetic field stronger than 2 kG on the star.

The magnetic field we derive here broadly agrees with previous measurements of the magnetic field through Zeeman broadening.  \citet{1995ApJ...439..939V} determine a magnetic field and filling factor from infrared lines, finding $B = 1445 \pm 58$ G and a filling factor $f = 0.088 \pm 0.008$.  This is similar to, but not formally consistent with our values, however they adopted a somewhat larger \logg\ (4.70) and \teff (5133 K) than we did.  In exploring the systematic impact of \logg\  and \teff\ they tested model with \logg$ = 4.55$ and  \teff$ = 4960$ K, finding $B = 1553$ G and $f = 0.096$, which agrees with our results within $3\sigma$ in $B$ and $2\sigma$ in $f$ (these differences being possibly linked to intrinsic magnetic variability).  A number of earlier magnetic estimates based on Zeeman broadening in the visible exist \citep[][e.g.]{Marcy1984, Gray1984, Saar1988, Mathys1989, Marcy1989}, but produce significantly larger values of $B$ or $f$ and \citet{1995ApJ...439..939V} question their reliability.  \citet{1997A&A...318..429R} provide a more consistent analysis of Zeeman broadening in the visible, although they prefer to report the product $Bf$ since they consider it less vulnerable to systematic errors (supported by the analysis of \citet{1995ApJ...439..939V} and we also find a possible negative covariance between the parameters).  \citet{1997A&A...318..429R} find $Bf = 165 \pm 30$ G, consistent with \citet{1995ApJ...439..939V} $Bf = 127 \pm 13$ G, and somewhat smaller than the product of the values from our two-component model $Bf = 237 \pm 36$ G, but within $2\sigma$ of our joint uncertainties.  From our multi-component model we calculate $\sum_{i} B_{i} f_{i} = 542 \pm 219$, although this quantity is dominated by the larger field bins with small, uncertain filling factors, thus we consider it effectively an upper limit. More recently \citet{Lehmann2015} investigated Zeeman broadening in \epseri\ spectra spanning 5 years, using a principal component analysis based approach.  They report $B\sqrt{f}$ values varying between $124 \pm 25$ and $230 \pm 21$ G, possibly varying with a 3 year cycle (e.g. \citealt{2013ApJ...763L..26M}).  They interpret this as a surface average magnetic field strength for a filling factor of 1, although for Zeeman broadening studies the filling factor is likely smaller, thus a direct comparison with our results is less obvious.  This quantity is consistent with our $Bf$, as well as \citet{1995ApJ...439..939V} and \citet{1997A&A...318..429R}, but is much smaller than our $B\sqrt{f}$ and that of \citet{1995ApJ...439..939V}.  While some discrepancies between studies are likely due to systematic errors or underestimated uncertainties, the results of \citet{Lehmann2015} suggest that there is also important temporal variability to the Zeeman broadening of \epseri. 


\section{CaII H and K emission}
\label{sec:sindex}

\begin{figure} 
\includegraphics[width=9cm]{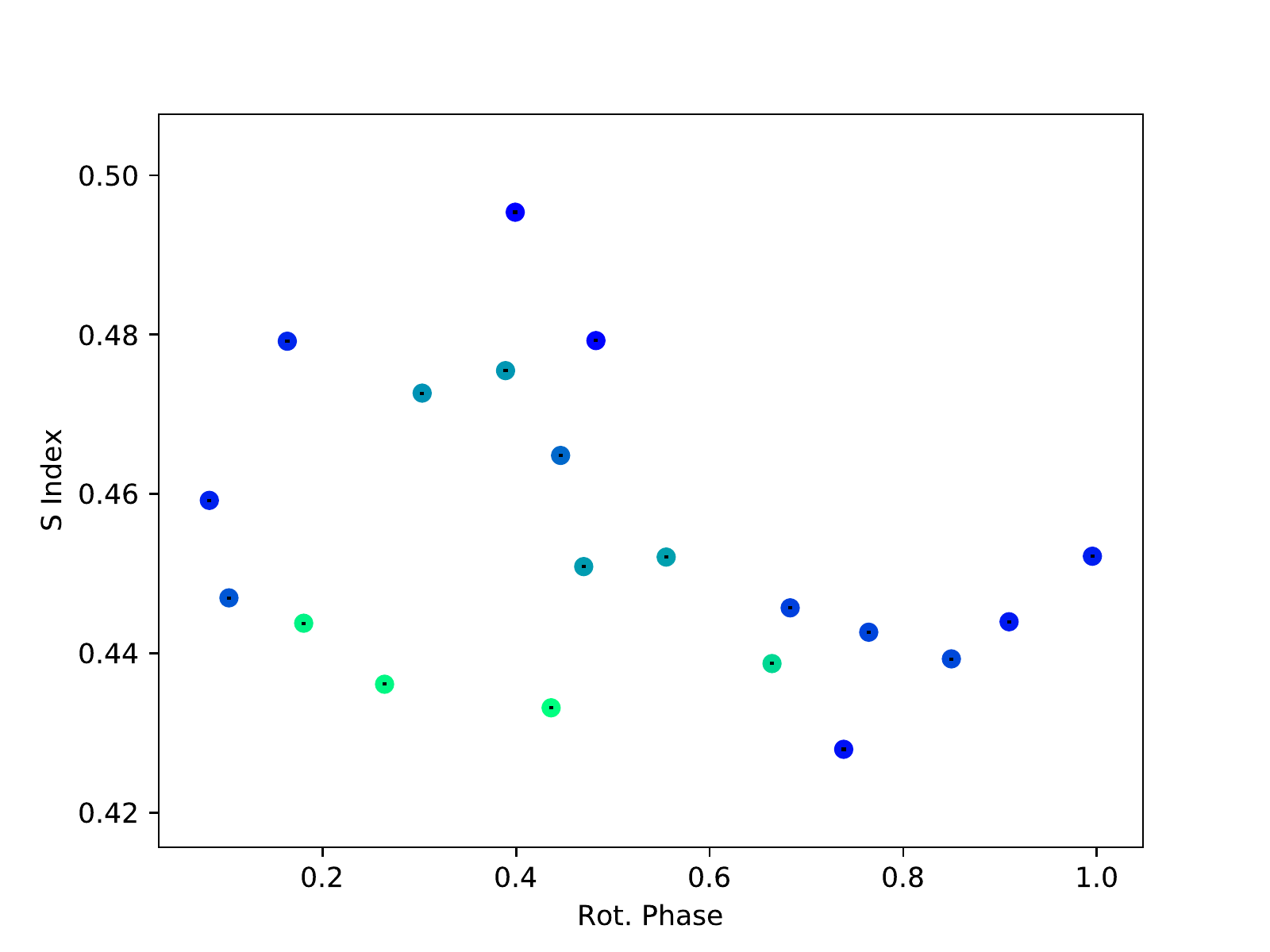} 
\caption{NARVAL S-index measurements, as a function of the rotational phase. Bluer (resp. greener) symbols show older (resp. more recent) observations.}
\label{fig:sindex}
\end{figure}

NARVAL spectra cover the CaII H \& K lines at 396.847 and 393.366 nm. Their core is seen in emission in all our set of \epseri\ spectra, as an effect of the sustained chromospheric activity. As a standard way to estimate the chromospheric emission and compare its value with archival measurements, we extract S-index values from our data following the guidelines of \cite{1984ApJ...279..763N}. This chromospheric indicator integrates the light flux in the cores of CaII H \& K using triangular bandpasses, and normalizes the core flux over two broader continuum bands (using a rectangular bandpass) taken on both sides of the doublet. Our measurements are calibrated according to \cite{2014MNRAS.444.3517M} to produce NARVAL S-index estimates that can be directly compared to Mount Wilson values. The resulting S-index time-series is reported in Tab. \ref{tab:journal}. The average value is close to 0.45, which is in the lower half of typical chromospheric emission measures for \epseri\, for which S-index values over the last five decades generally range from 0.4 to 0.6 \citep{2013ApJ...763L..26M}. Our values are in overall agreement with the contemporaneous monitoring of \cite{2020A&A...636A..49C}, who report an unexpected drop in chromospheric activity with respect to the previously regular S-index fluctuations (so that our set of observations, that was expected to be close to an S-index maximum, are in fact typical of a lower activity state). 

Statistical error bars can be computed for every individual estimate but these formal uncertainties are, in practice, dominated by residual inaccuracies in continuum normalisation around the calcium doublet (even if, by definition, the S-index corrects a fair part of normalization issues). To obtain an empirical estimate of typical error bars, we consider the dispersion of S-index measurements for Pollux, a quiet red giant with chromospheric emission close to the basal level and a K0 spectral type relatively close to the spectral type of \epseri. From a series of 266 observations performed with NARVAL and ESPaDOnS over several years, at a \sn\ close to the one achieved here, we measure a standard deviation of the S-index of $\sim 0.002$ \citep{auriere21}. This value is taken as a proxy of the S-index uncertainty for \epseri. 

The S-index variations are plotted, against rotational phases, in Fig. \ref{fig:sindex}. The phase dependence is not as prominent as the one recorded for the longitudinal field, to the point that the rotation period cannot be unambiguously identified using a GPR model similar to the one described in Sec. \ref{sec:bl}, or using a simple Lomb-Scargle approach. Fast evolution of chromospheric structures may be responsible for the relatively large scatter, with non-rotational evolution especially obvious between phases 0.2 and 0.4, where S-index levels have sharply dropped during the time span of observations. In the phase range $[0.16,0.18]$, we record a decrease from $S \approx 0.48$ at $HJD = 2458389.6$ to $S \approx 0.44$ at $HJD = 2458436.5$, which is well above uncertainties. During a comparable time interval, a more limited drop is observed at phases 0.67-0.68, with S-index values decreasing from 0.45 to 0.44. We therefore observe a sudden transition from a marked rotational modulation of chromospheric emission to a flatter one, with most of the recorded evolution taking place during the last three days of the monitoring. Prior to this specific event, the rotational modulation is more visible, with the strongest emission around phase 0.4 and the weakest chromospheric flux around phase 0.7. Even during this first part of the time series, a significant scatter is present and may be a hint that frequent transitory events contribute to the S-index. 

We also estimated an $H_\alpha$ index, following the methodology of \cite{2002AJ....123.3356G}, and obtain a similar trend, although the contrast between high and low emission is not as large using this specific chromospheric tracer (see Fig. \ref{fig:all}). We finally checked the chromospheric emission in the CaII~IRT triplet, using the index defined by \cite{2013LNP...857..231P}. In this last case (not shown here), neither the early rotational modulation or the late drop are visible in the time-series.

\section{Radial velocities}
\label{sec:rv}

Radial velocities have been estimated for both sets of SPIRou and NARVAL LSD profiles, with values reported in Tab. \ref{tab:journal}. The accuracy of NARVAL radial velocity estimates is mainly limited by the wavelength calibration using telluric lines, as part of the standard reduction process. From observations of $\tau$~Boo, \cite{2007A&A...473..651M} reported a precision of the order of 30~\ms. Uncertainties for SPIRou are taken equal to the standard deviation of the values obtained for a given night, and range between 1 and 2 \ms. SPIRou values tend to be smaller than NARVAL values. Although the difference stays mostly within uncertainties, a different absolute calibration between the two instrument is likely causing this offset.   

All available values are plotted against the rotation phase in Fig. \ref{fig:all}. The rotational signal is not detected in the NARVAL time-series, which is in agreement with a peak-to-peak activity jitter of about 30~\ms\ reported by \cite{Giguere16}, below our detection threshold. SPIRou measurements are too sparse to reveal any rotational dependence. Variations observed over the four nights reach 24~\ms peak-to-peak, which remains at a level that we should not expect to detect with NARVAL. 

\section{Brightness fluctuations}
\label{sec:bright}

The temporal variability of the optical lightcurve across the TESS observing window is dominated by quasi-periodic, smooth variations that, to the naked eye, look roughly consistent with the known rotation period of 11.68~d. The $\sim 2$ mmag amplitude is smaller than the $\sim 6$ mmag reported by \cite{Giguere16}, suggesting that our observing epoch was characterized either by a smaller number of spots and plages, or by a more axisymmetric distribution of surface brightness inhomogeneities.  We model the photometric variations through a GPR model, based on a pseudo-periodic type of GP and an MCMC exploration of the hyperparameter space similar to the one described in Sec. \ref{sec:bl}. 

With this aim, we first reduce the number of data points by rebinning the light curve, as the temporal sampling offered by TESS exceeds by far the number of observations required to sample the rotation period of \epseri. In each bin, the average is weighted with the inverse square of the error bars, for both time and flux, and the error is the weighted standard deviation, as described by the following equations.

\begin{equation}
<{\rm BJD}>_{\rm bin} = \dfrac{\sum_{{\rm BJD}_i \in {\rm bin}}\dfrac{{\rm BJD}_i}{\sigma_i^2}}{\sum_{{\rm BJD}_i \in {\rm bin}}\dfrac{1}{\sigma_i^2}}\\
\end{equation}

\begin{equation}
<F>_{\rm bin} = \dfrac{\sum_{{\rm BJD}_i \in {\rm bin}}\dfrac{F_i}{\sigma_i^2}}{\sum_{{\rm BJD}_i \in {\rm bin}}\dfrac{1}{\sigma_i^2}}\\
\end{equation}

\begin{equation}
\sigma_{<F>,{\rm bin}} = \sqrt{\dfrac{\sum_{{\rm BJD}_i \in {\rm bin}}{\dfrac{(F_i-<F>_{\rm bin})^2}{\sigma_i^2}}}{\sum_{{\rm BJD}_i \in {\rm bin}}\dfrac{1}{\sigma_i^2}}}
\end{equation}

\noindent where BJD is the barycentric Julian date, $F$ is the flux, and $\sigma_i$ is the error bar on each photometric measurement. We chose to represent the local error $\sigma_{<F>,{\rm bin}}$ by the standard deviation because the observed dispersion is around 5 times higher than the photon noise level. We generated one light curve with 0.2~d bins (114 points in total), and a second one with 0.05~d bins (456 points).  

Using 0.2~d bins, the most likely value of the GP hyperparameters include a rotation period of 11.62~$\pm$~0.15~d (best: 11.63~d) and a decay time of 43$^{+18}_{-13}$~d (best: 42~d). Using 0.05~d bins, the same procedure leads to a rotation period of 11.56$~\pm$~0.19~d (best: 11.42~d) and a decay time equal to 33$^{+13}_{-9}$~d (best: 35~d), in agreement (within uncertainties) with the values obtained using 0.2~d bins. 

While intrinsic evolution of the spot pattern is likely to dominate the observed non-rotational variability of the light curve, a few transitory events visible in Fig. \ref{fig:TESS} (e.g. at Julian date $\sim 2458417$) may also contribute to reduce the decay time. This may be reflected in the slightly smaller decay time derived with 0.05~d bins, since short-term fluctuations are better filtered out using 0.2~d bins.

\section{Tomographic mapping of the large-scale surface magnetic field}
\label{sec:zdi}

\begin{figure} 
\includegraphics[width=7cm]{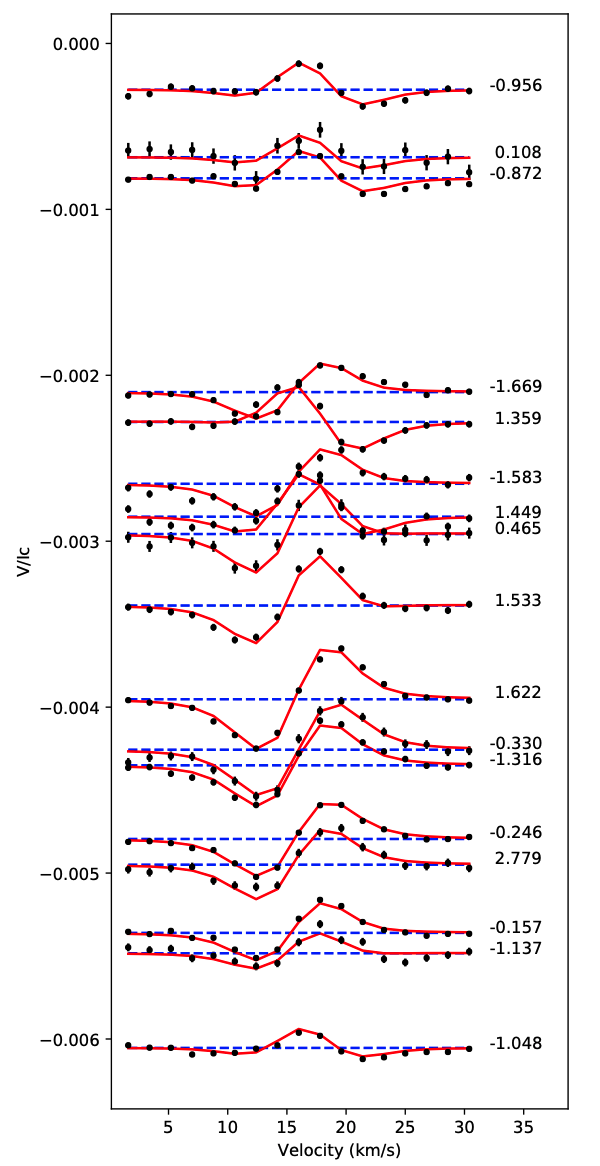} 
\caption{Time-series of NARVAL Stokes $V$ LSD profiles (black dots), and synthetic profiles produced by the ZDI model (red lines). Successive profiles are translated vertically for display purposes, with vertical shifts proportional to phase gaps. The rotation cycle is indicated on the right part of the plot. Dashed blue horizontal lines depict the zero level of each profiles.}
\label{fig:stokesv}
\end{figure}

The series of NARVAL Stokes $V$ pseudo-profiles was used to model the large-scale magnetic field geometry by means of ZDI. This tomographic approach was first proposed by \cite{semel89} and is based on a maximum entropy regularisation of the ill-posed problem of inverting a set of circularly polarised Zeeman signatures. More specifically, we use here a magnetic model where the surface magnetic field is decomposed over a spherical harmonics frame \citep{donati06}, through the Python implementation of \cite{2018MNRAS.474.4956F}. We also assume that the local Stokes $I$ line profiles (associated to each surface pixel of our model) take the shape of a Voigt profile, following \cite{2008MNRAS.390..567M} and \cite{2018MNRAS.481.5286F}. 

The input projected rotational velocity and the inclination angle are taken equal to the values selected by \cite{2014A&A...569A..79J,2017MNRAS.471L..96J}, with \vsin$=2.59$~\kms\ (see Sec. \ref{sec:param_SPIRou}) and $i=46$\degr{}. The \drot{} parameters were refined compared to the earlier estimate of \cite{2014A&A...569A..79J}, but this specific point will be discussed in Sec. \ref{sec:drot}. The set of observed Stokes $V$ LSD profiles, as well as the set of synthetic profiles produced by the ZDI model, are plotted in Fig. \ref{fig:stokesv}. 

\begin{figure} 
\includegraphics[width=\linewidth]{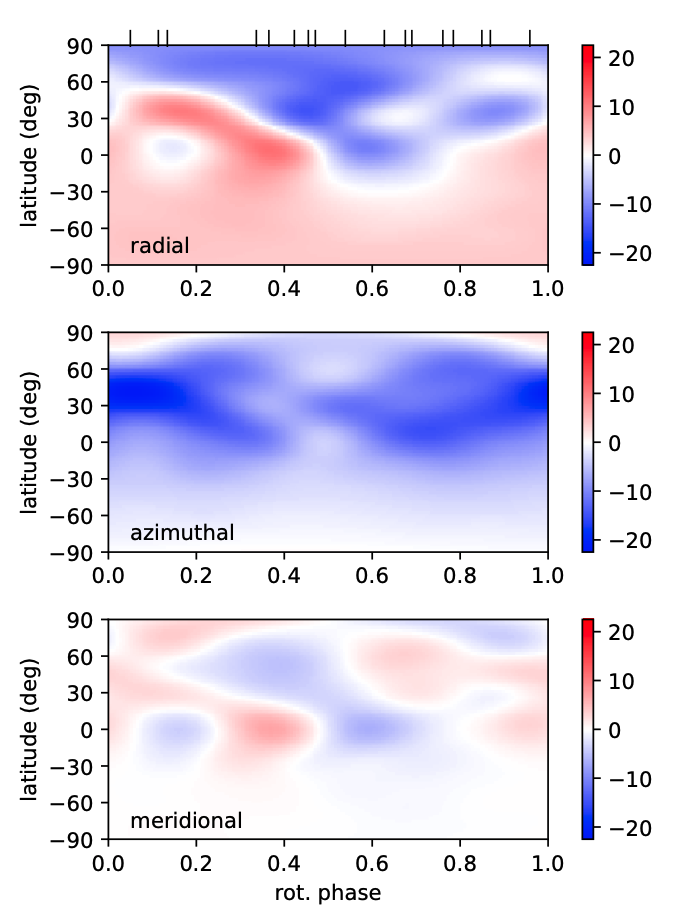} 
\caption{ZDI reconstruction of the large-scale surface magnetic geometry of \epseri{}, using NARVAL data. Every chart displays a different component of the field in spherical coordinates, color coded according to the field strength (expressed in gauss). The vertical ticks on top of the radial field map show the rotational phases of observations.}
\label{fig:map}
\end{figure}

The magnetic map of Fig. \ref{fig:map} (including differential rotation) highlights a complex distribution of magnetic regions, although the small \vsin{} value limits the spatial resolution of the ZDI inversion. The average field strength is equal to 9.2~G, while the peak field modulus reaches 20~G. The toroidal magnetic component hosts most of the surface magnetic energy (68\%). The field geometry is also predominantly axi-symmetric, with as much as $73\%$ of the magnetic energy in spherical harmonics modes with $m = 0$.
A majority of the magnetic energy is obtained in the lowest-order components of the spherical harmonics expansion, with $92\%$ in modes with $1 \leq \ell \leq 3$. If we consider the poloidal field component alone, we find $34\%$ of its magnetic energy in the dipole, $28\%$ in the quadrupole and $17\%$ in the octupole. 

Comparing these general field properties to values previously published by \cite{2014A&A...569A..79J} and \cite{2017MNRAS.471L..96J} is hampered by the different local profile shape used in our study (Voigt versus gaussian), as well as a slightly different \vsin\ value (2.2~\kms\ in these previous studies). To allow for a more direct comparison, we reconstructed a magnetic model based on these alternate ZDI settings. The fit to Stokes $I$ profiles is significantly better with a Voigt profile that allows for a good adjustment of both the core and wings of LSD pseudo-profiles, while a gaussian profile provides only a good fit to the line core. The consequence on the Stokes $V$ adjustment is, in practice, less dramatic with general reconstructed field characteristics reasonably close to the ones found using a Voigt profile. The alternate model leads to a mean field strength of 8~G, 64\% of the magnetic energy in the toroidal component, or 68\% of the energy in axisymmetric modes. The average field value obtained here is on the lower end of previously published values (obtained close to activity minimum). On the other hand, the fraction of energy in the toroidal component is the highest reported to date. As noticed by \cite{2015MNRAS.453.4301S}, higher toroidal fields generally correspond to higher field axisymmetry, which is also observed here with a magnetic geometry among the most axisymmetric recorded so far for \epseri. This atypical combination of magnetic properties may suggest that the dynamo activity of \epseri{}, which was already reported in the past to change from chaotic to cyclic \citep{2013ApJ...763L..26M} may have entered another new phase, as suggested by the absence of the expected activity maximum in early 2019 \citep{2020A&A...636A..49C} and low flaring activity in radio observations obtained in 2019 by \cite{2020arXiv201005929S}.   

\section{Surface differential rotation}
\label{sec:drot}

Reconstructing the magnetic field geometry of \epseri\ under the simple assumption of solid-body rotation leads to a disappointing reduced \kis\ (\kisr{}\ hereafter) equal to 3.2. We try here to improve the model by assuming that the surface experiences a latitudinal shear.

The ZDI model includes the possibility for the surface field geometry to change with time, under the progressive shear imposed by a solar-like differential rotation law described by the following equation:

\begin{equation}
\Omega (l) = \Omega_{\rm eq} - d\Omega.\sin^2(l)
\label{eq:drot}
\end{equation}ı

Here $\Omega (l)$ is the rotation rate at stellar latitude $l$, $\Omega_{\rm eq}$ the rotation rate at equatorial latitude, and $d\Omega$ stands for the difference in rotation rate between the equator and pole. Following \cite{2000MNRAS.316..699D} and \cite{petit02}, the two free parameters of this simple law are estimated by running a large number of ZDI models over a grid of values of the two parameters ($\Omega_{\rm eq}; d\Omega$), searching for values that optimize the ZDI model ({\em i.e.} that minimize the \kisr\ of the model, at fixed entropy value).

\begin{figure} 
\includegraphics[width=9cm]{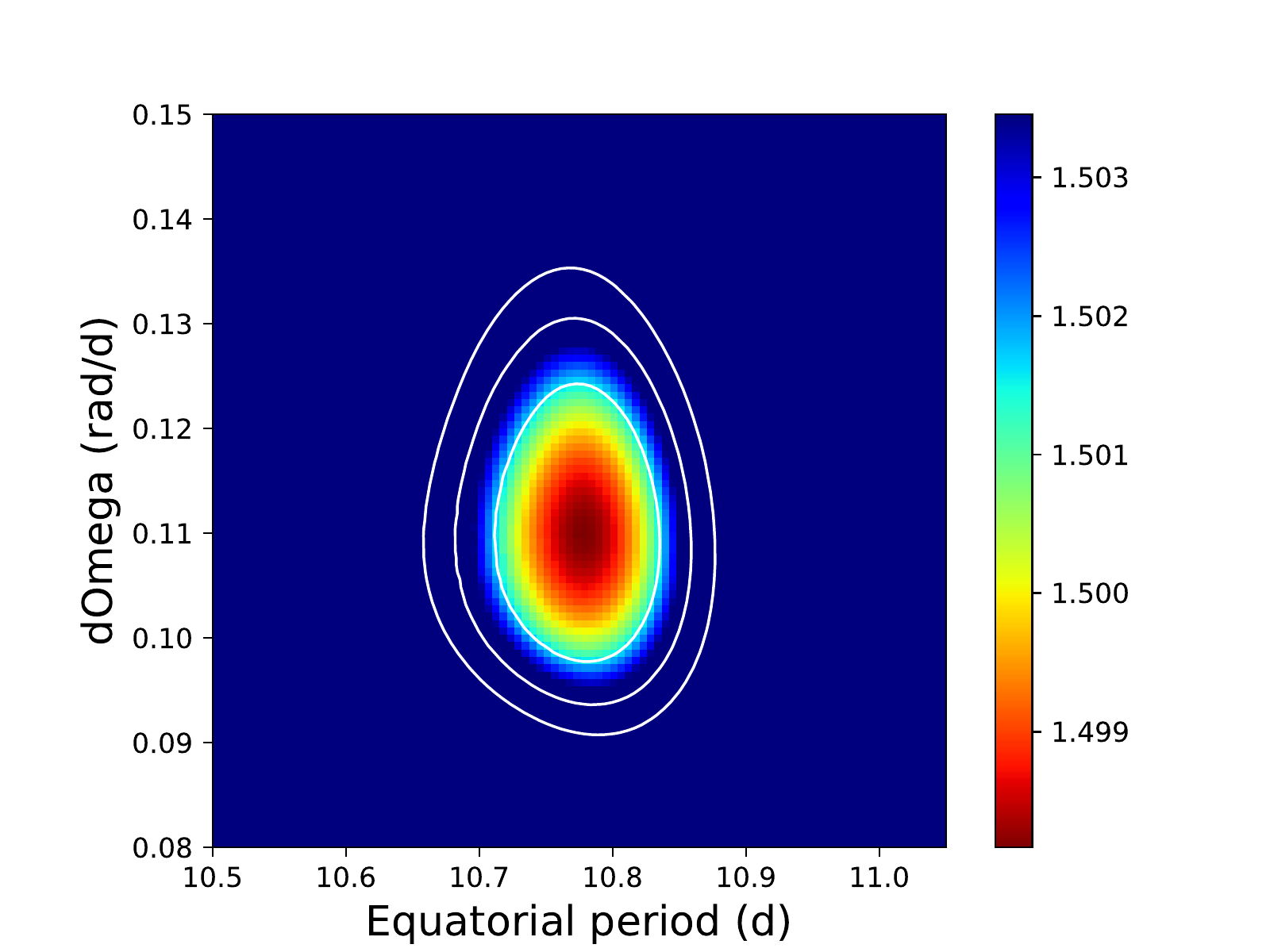} 
\caption{Reduced $\chi^2$ landscape obtained for a grid of ZDI models implementing Eq. \ref{eq:drot}. The three concentric, white contours depict the $1\sigma$, $2\sigma$ and $3\sigma$ limits around the \kisr\ minimum.}
\label{fig:drot}
\end{figure}

A clear minimum is found in the $\chi^2$ landscape (Fig. \ref{fig:drot}), for an equatorial rotation period $P_{\rm eq} = 2\pi/\Omega_{\rm eq}= 10.77 \pm 0.06$~d and $d \Omega = 0.11 \pm 0.01$~\rpd{}. The resulting \kisr\ value is close to 1.5, unveiling a much better fit to the data when the surface is assumed to be sheared by differential rotation. The fact that our best \kisr\ is still larger than one is indicative that other phenomena (e.g. continuous emergence and decay of magnetic spots) contribute as well to the magnetic evolution. 

The periods derived from longitudinal field measurements and photometry (Sec. \ref{sec:bl} and \ref{sec:bright}) are longer than the equatorial period obtained through the ZDI model. They are also shorter than the rotation period we can extrapolate at polar latitude from Eq. \ref{eq:drot} (about 13.3~d). This is expected if the main surface features contributing to the longitudinal magnetic field or to the photometric variability are located at intermediate latitudes. The shear level obtained here is about twice the solar value, and leads to a laptime (the time it takes for the equator to be one rotation cycle ahead of the pole) of  $2\pi/d \Omega = 57 \pm 5$~d, providing us with a third typical timescale of surface intrinsic evolution.

\section{Discussion}
\label{sec:discussion}

\subsection{Phase dependence of activity tracers}

The different tracers investigated here of the magnetic activity at photospheric and chromospheric levels produce a diversity of temporal signatures, primarily highlighted by their different rotational dependence. In absolute value, the longitudinal magnetic field is maximal around phase $0.6$, which on the magnetic map is translated as a peak in the radial field strength (with negative polarity). In the light curve, this specific rotation phase is off any extrema, with a brightness maximum at phase 0.75, while the minimum in the light flux is recorded around phase 0.45. 

The longitudinal field is switching from positive to negative at phase 0.4, close to the minimum light flux recorded by TESS. On the magnetic map, this phase is mostly characterized by a minimum in the azimuthal field, as well as a transition from a positive to a negative radial field polarity at intermediate and low latitudes (which may suggest that the magnetic equator preferentially hosts cooler spots than the magnetic pole). Another zero-crossing of the longitudinal field is recorded around phases 0.95-1, where the radial field at intermediate latitudes on the ZDI map goes from a negative to a positive polarity. This second field minimum has no remarkable counterpart on the light curve, which stays close to its average value in the same phase interval. 

Prior to its weakening during the last days of the NARVAL time-series, the S-index of \epseri\ was maximal around phase 0.4, where in November the light curve was near its minimum, and where the radial component of the large-scale magnetic field was switching polarity. During the last rotation period covered in the time-series, the phase dependence of the S-index seems to become flatter, although late observations of phases above 0.7 are missing to track this fast evolution throughout the whole rotation cycle. We note that this sharp evolution, taking place within a few days, is not reflected in the longitudinal field values, for which the phase modulation is mostly stable over the whole time-series. The TESS data, which are representative of the second half of the time-span covered by NARVAL, display a same peak-to-peak amplitude (about 5 mmag) in the first and second observed rotation period (blue and orange symbols in the upper panel of Fig. \ref{fig:all}). The incomplete phase coverage of the first period (with no available data for phases above 0.65, missing the phase of maximal flux) leads to this apparent stability, while the photometric amplitude seems to have mostly decreased between the two periods, at least for the brightest phases (phases 0.0 to 0.35, and phases above 0.6), while the dimmest phases seem to be unaffected by this evolution. 

Among the possible explanations of the apparently discrepant phase dependence of the different magnetic tracers, one obviously at play is their different spatial resolution. Magnetic fields on \epseri\ are likely distributed in a complex pattern featuring both field polarities, and nearly identical radial velocities (due to the relatively small \vsin). The visible magnetic spots have most of their Stokes $V$ signatures mutually canceled, and the remaining signal is limited to the largest spatial scales of the surface field. Zeeman broadening, on the opposite, does not depend on field polarity, explaining the much weaker field strength reconstructed in the ZDI map, compared to the Zeeman broadening of nIR lines (a detailed discussion of this aspect can be found in \citealt{2019ApJ...876..118S}). The absence of any detectable phase dependence of the Zeeman broadening tends to support the picture of complex distribution of magnetic regions. The broad band stellar photometry is a degenerate observation, with a disc-integrated brightness reflecting the surface balance between dark and bright surface features. Finally, the chromospheric emission is a cumulative effect, unaffected by the local orientation of the chromospheric magnetic field. 

Another effect contributing to a different phase dependence between the different quantities investigated here is their different limb visibility. While polarized Zeeman signatures produced by spots with radial magnetic fields are best seen close to disc center, azimuthal magnetic fields will have larger Stokes $V$ amplitudes at intermediate limb angles \citep{1997A&A...326.1135D}. Similar considerations impact the interpretation of the TESS light curve, with the contribution of dark spots being maximal close to disc center while, by analogy with the Sun, faculae may be brighter close to the limb \citep{1979SoPh...63..251H}, which may also affect the phase dependence of the S-index.   

\subsection{Characteristic timescales for short-term surface evolution}

Longitudinal field measurements, photometric variations and the differential rotation model provide us with three independent estimates of evolution timescales. Two decay times are obtained from GPR applied to $B_{\rm eff}$ and TESS data, while $2\pi / d\Omega$ gives a characteristic timescale of the surface shear. All three estimates are consistent within uncertainties. The longest one is the differential rotation laptime ($57 \pm 5$~d). The decay time deduced from photometry comes second and is equal to 43$^{+18}_{-13}$~d, while the same quantity estimated from $B_{\rm eff}$ measurements is the shortest and is equal to $31^{+9}_{-7}$~d. The laptime is linked to the specific component of the surface evolution driven globally by differential rotation, while the two other estimates include as well the contribution of the limited lifetime of surface structures (note that the laptime estimated for differential rotation can also be biased whenever the shear tracers come and go, see \citealt{petit02}). We interpret this difference as the cause of the shorter timescales obtained out of the light curve and longitudinal field data.   

\section{Conclusions and prospects}

This multi-instrumental view of \epseri\ reveals how different tracers of magnetism and activity carry different and complementary information about the surface activity linked to the vivid dynamo action of young solar-like stars. Each available measurement brings its own set of clues about the underlying emergence and decay of active regions, each specific tracer being limited by its own degeneracy, spatial resolution, temporal resolution, or limb dependence. The conclusions that we can draw from the diverse data presented here are also limited by the non-simultaneity of the observations. This would advocate the future development of high-resolution spectropolarimeters covering both the optical and near-infrared domain, as would be offered by a combination of SPIRou and ESPaDOnS. In addition to help reach a better understanding of photospheric and chromospheric stellar activity, such instruments would help progress in the filtering of stellar activity, as part of RV exoplanet search and characterization around cool active stars. 

\begin{acknowledgements}

JFD acknowledges funding from from the European Research Council (ERC) under the H2020 research \& innovation programme (grant agreement \# 740651 NewWorlds). AAV acknowledges funding from the European Research Council (ERC) under the European Union's Horizon 2020 research and innovation programme (grant agreement No 817540, ASTROFLOW). This research made use of the SIMBAD database operated at CDS, Strasbourg, France, and the NASA's Astrophysics Data System Abstract Service.

\end{acknowledgements}

\bibliographystyle{aa}
\bibliography{epseri.bib}

\begin{thebibliography}{77}
\expandafter\ifx\csname natexlab\endcsname\relax\def\natexlab#1{#1}\fi

\bibitem[{{Anglada-Escud{\'e}} \& {Butler}(2012)}]{2012ApJS..200...15A}
{Anglada-Escud{\'e}}, G. \& {Butler}, R.~P. 2012, \apjs, 200, 15

\bibitem[{{Artigau} {et~al.}(2014){Artigau}, {Astudillo-Defru}, {Delfosse},
  {Bouchy}, {Bonfils}, {Lovis}, {Pepe}, {Moutou}, {Donati}, {Doyon}, \&
  {Malo}}]{2014SPIE.9149E..05A}
{Artigau}, {\'E}., {Astudillo-Defru}, N., {Delfosse}, X., {et~al.} 2014,
  Society of Photo-Optical Instrumentation Engineers (SPIE) Conference Series,
  Vol. 9149, {Telluric-line subtraction in high-accuracy velocimetry: a
  PCA-based approach}, 914905

\bibitem[{{Auri{\`e}re}(2003)}]{auriere03}
{Auri{\`e}re}, M. 2003, in EAS Publications Series, Vol.~9, EAS Publications
  Series, ed. J.~{Arnaud} \& N.~{Meunier}, 105

\bibitem[{{Auri{\`e}re} {et~al.}(2021){Auri{\`e}re}, {Petit}, {Mathias},
  {Konstantinova-Antova}, {Charbonnel}, {Donati}, {Espagnet}, {Folsom},
  {Roudier}, \& {Wade}}]{auriere21}
{Auri{\`e}re}, M., {Petit}, P., {Mathias}, P., {et~al.} 2021, arXiv e-prints,
  arXiv:2101.02016

\bibitem[{{Baines} \& {Armstrong}(2012)}]{2012ApJ...744..138B}
{Baines}, E.~K. \& {Armstrong}, J.~T. 2012, \apj, 744, 138

\bibitem[{{Baliunas} {et~al.}(1995){Baliunas}, {Donahue}, {Soon}, {Horne},
  {Frazer}, {Woodard-Eklund}, {Bradford}, {Rao}, {Wilson}, {Zhang}, {Bennett},
  {Briggs}, {Carroll}, {Duncan}, {Figueroa}, {Lanning}, {Misch}, {Mueller},
  {Noyes}, {Poppe}, {Porter}, {Robinson}, {Russell}, {Shelton}, {Soyumer},
  {Vaughan}, \& {Whitney}}]{1995ApJ...438..269B}
{Baliunas}, S.~L., {Donahue}, R.~A., {Soon}, W.~H., {et~al.} 1995, \apj, 438,
  269

\bibitem[{{Barnes}(2007)}]{2007ApJ...669.1167B}
{Barnes}, S.~A. 2007, \apj, 669, 1167

\bibitem[{{Berdyugina}(2005)}]{2005LRSP....2....8B}
{Berdyugina}, S.~V. 2005, Living Reviews in Solar Physics, 2, 8

\bibitem[{{Booth} {et~al.}(2017){Booth}, {Dent}, {Jord{\'a}n}, {Lestrade},
  {Hales}, {Wyatt}, {Casassus}, {Ertel}, {Greaves}, {Kennedy}, {Matr{\`a}},
  {Augereau}, \& {Villard}}]{2017MNRAS.469.3200B}
{Booth}, M., {Dent}, W. R.~F., {Jord{\'a}n}, A., {et~al.} 2017, \mnras, 469,
  3200

\bibitem[{{Boro Saikia} {et~al.}(2018){Boro Saikia}, {Marvin}, {Jeffers},
  {Reiners}, {Cameron}, {Marsden}, {Petit}, {Warnecke}, \&
  {Yadav}}]{2018A&A...616A.108B}
{Boro Saikia}, S., {Marvin}, C.~J., {Jeffers}, S.~V., {et~al.} 2018, \aap, 616,
  A108

\bibitem[{{Coffaro} {et~al.}(2020){Coffaro}, {Stelzer}, {Orlando}, {Hall},
  {Metcalfe}, {Wolter}, {Mittag}, {Sanz-Forcada}, {Schneider}, \&
  {Ducci}}]{2020A&A...636A..49C}
{Coffaro}, M., {Stelzer}, B., {Orlando}, S., {et~al.} 2020, \aap, 636, A49

\bibitem[{{di Folco} {et~al.}(2007){di Folco}, {Absil}, {Augereau}, {M{\'e}rand
  }, {Coud{\'e} du Foresto}, {Th{\'e}venin}, {Defr{\`e}re}, {Kervella}, {ten
  Brummelaar}, {McAlister}, {Ridgway}, {Sturmann}, {Sturmann}, \&
  {Turner}}]{diFolco2007}
{di Folco}, E., {Absil}, O., {Augereau}, J.~C., {et~al.} 2007, \aap, 475, 243

\bibitem[{{do Nascimento} {et~al.}(2016){do Nascimento}, {Vidotto}, {Petit},
  {Folsom}, {Castro}, {Marsden}, {Morin}, {Porto de Mello}, {Meibom},
  {Jeffers}, {Guinan}, \& {Ribas}}]{2016ApJ...820L..15D}
{do Nascimento}, J.~D., J., {Vidotto}, A.~A., {Petit}, P., {et~al.} 2016,
  \apjl, 820, L15

\bibitem[{{Donahue} {et~al.}(1996){Donahue}, {Saar}, \&
  {Baliunas}}]{1996ApJ...466..384D}
{Donahue}, R.~A., {Saar}, S.~H., \& {Baliunas}, S.~L. 1996, \apj, 466, 384

\bibitem[{{Donati} \& {Brown}(1997)}]{1997A&A...326.1135D}
{Donati}, J.~F. \& {Brown}, S.~F. 1997, \aap, 326, 1135

\bibitem[{{Donati} {et~al.}(2006){Donati}, {Howarth}, {Jardine}, {Petit},
  {Catala}, {Landstreet}, {Bouret}, {Alecian}, {Barnes}, {Forveille},
  {Paletou}, \& {Manset}}]{donati06}
{Donati}, J.-F., {Howarth}, I.~D., {Jardine}, M.~M., {et~al.} 2006, \mnras,
  370, 629

\bibitem[{{Donati} {et~al.}(2020){Donati}, {Kouach}, {Moutou}, {Doyon},
  {Delfosse}, {Artigau}, {Baratchart}, {Lacombe}, {Barrick}, {H{\'e}brard},
  {Bouchy}, {Saddlemyer}, {Par{\'e}s}, {Rabou}, {Micheau}, {Dolon}, {Reshetov},
  {Challita}, {Carmona}, {Striebig}, {Thibault}, {Martioli}, {Cook},
  {Fouqu{\'e}}, {Vermeulen}, {Wang}, {Arnold}, {Pepe}, {Boisse}, {Figueira},
  {Bouvier}, {Ray}, {Feugeade}, {Morin}, {Alencar}, {Hobson}, {Castilho},
  {Udry}, {Santos}, {Hernandez}, {Benedict}, {Vall{\'e}e}, {Gallou}, {Dupieux},
  {Larrieu}, {Perruchot}, {Sottile}, {Moreau}, {Usher}, {Baril}, {Wildi},
  {Chazelas}, {Malo}, {Bonfils}, {Loop}, {Kerley}, {Wevers}, {Dunn}, {Pazder},
  {Macdonald}, {Dubois}, {Carri{\'e}}, {Valentin}, {Henault}, {Yan}, \&
  {Steinmetz}}]{2020MNRAS.tmp.2502D}
{Donati}, J.~F., {Kouach}, D., {Moutou}, C., {et~al.} 2020, \mnras
  [\eprint[arXiv]{2008.08949}]

\bibitem[{{Donati} {et~al.}(2000){Donati}, {Mengel}, {Carter}, {Marsden},
  {Collier Cameron}, \& {Wichmann}}]{2000MNRAS.316..699D}
{Donati}, J.~F., {Mengel}, M., {Carter}, B.~D., {et~al.} 2000, \mnras, 316, 699

\bibitem[{{Donati} {et~al.}(1997){Donati}, {Semel}, {Carter}, {Rees}, \&
  {Collier Cameron}}]{donati97}
{Donati}, J.-F., {Semel}, M., {Carter}, B.~D., {Rees}, D.~E., \& {Collier
  Cameron}, A. 1997, \mnras, 291, 658

\bibitem[{{Donati} {et~al.}(1990){Donati}, {Semel}, {Rees}, {Taylor}, \&
  {Robinson}}]{1990A&A...232L...1D}
{Donati}, J.~F., {Semel}, M., {Rees}, D.~E., {Taylor}, K., \& {Robinson}, R.~D.
  1990, \aap, 232, L1

\bibitem[{{Folsom} {et~al.}(2018{\natexlab{a}}){Folsom}, {Bouvier}, {Petit},
  {L{\`e}bre}, {Amard}, {Palacios}, {Morin}, {Donati}, \&
  {Vidotto}}]{2018MNRAS.474.4956F}
{Folsom}, C.~P., {Bouvier}, J., {Petit}, P., {et~al.} 2018{\natexlab{a}},
  \mnras, 474, 4956

\bibitem[{{Folsom} {et~al.}(2018{\natexlab{b}}){Folsom}, {Fossati}, {Wood},
  {Sreejith}, {Cubillos}, {Vidotto}, {Alecian}, {Girish}, {Lichtenegger},
  {Murthy}, {Petit}, \& {Valyavin}}]{2018MNRAS.481.5286F}
{Folsom}, C.~P., {Fossati}, L., {Wood}, B.~E., {et~al.} 2018{\natexlab{b}},
  \mnras, 481, 5286

\bibitem[{{Folsom} {et~al.}(2016){Folsom}, {Petit}, {Bouvier}, {L{\`e}bre},
  {Amard}, {Palacios}, {Morin}, {Donati}, {Jeffers}, {Marsden}, \&
  {Vidotto}}]{2016MNRAS.457..580F}
{Folsom}, C.~P., {Petit}, P., {Bouvier}, J., {et~al.} 2016, \mnras, 457, 580

\bibitem[{{Gaia Collaboration} {et~al.}(2018){Gaia Collaboration}, {Brown},
  {Vallenari}, {Prusti}, {de Bruijne}, {Babusiaux}, {Bailer-Jones}, {Biermann},
  {Evans}, {Eyer}, {Jansen}, {Jordi}, {Klioner}, {Lammers}, {Lindegren},
  {Luri}, {Mignard}, {Panem}, {Pourbaix}, {Randich}, {Sartoretti}, {Siddiqui},
  {Soubiran}, {van Leeuwen}, {Walton}, {Arenou}, {Bastian}, {Cropper},
  {Drimmel}, {Katz}, {Lattanzi}, {Bakker}, {Cacciari}, {Casta{\~n}eda},
  {Chaoul}, {Cheek}, {De Angeli}, {Fabricius}, {Guerra}, {Holl}, {Masana},
  {Messineo}, {Mowlavi}, {Nienartowicz}, {Panuzzo}, {Portell}, {Riello},
  {Seabroke}, {Tanga}, {Th{\'e}venin}, {Gracia-Abril}, {Comoretto},
  {Garcia-Reinaldos}, {Teyssier}, {Altmann}, {Andrae}, {Audard},
  {Bellas-Velidis}, {Benson}, {Berthier}, {Blomme}, {Burgess}, {Busso},
  {Carry}, {Cellino}, {Clementini}, {Clotet}, {Creevey}, {Davidson}, {De
  Ridder}, {Delchambre}, {Dell'Oro}, {Ducourant},
  {Fern{\'a}ndez-Hern{\'a}ndez}, {Fouesneau}, {Fr{\'e}mat}, {Galluccio},
  {Garc{\'\i}a-Torres}, {Gonz{\'a}lez-N{\'u}{\~n}ez}, {Gonz{\'a}lez-Vidal},
  {Gosset}, {Guy}, {Halbwachs}, {Hambly}, {Harrison}, {Hern{\'a}ndez},
  {Hestroffer}, {Hodgkin}, {Hutton}, {Jasniewicz}, {Jean-Antoine-Piccolo},
  {Jordan}, {Korn}, {Krone-Martins}, {Lanzafame}, {Lebzelter}, {L{\"o}ffler},
  {Manteiga}, {Marrese}, {Mart{\'\i}n-Fleitas}, {Moitinho}, {Mora}, {Muinonen},
  {Osinde}, {Pancino}, {Pauwels}, {Petit}, {Recio-Blanco}, {Richards},
  {Rimoldini}, {Robin}, {Sarro}, {Siopis}, {Smith}, {Sozzetti}, {S{\"u}veges},
  {Torra}, {van Reeven}, {Abbas}, {Abreu Aramburu}, {Accart}, {Aerts},
  {Altavilla}, {{\'A}lvarez}, {Alvarez}, {Alves}, {Anderson}, {Andrei},
  {Anglada Varela}, {Antiche}, {Antoja}, {Arcay}, {Astraatmadja}, {Bach},
  {Baker}, {Balaguer-N{\'u}{\~n}ez}, {Balm}, {Barache}, {Barata}, {Barbato},
  {Barblan}, {Barklem}, {Barrado}, {Barros}, {Barstow}, {Bartholom{\'e}
  Mu{\~n}oz}, {Bassilana}, {Becciani}, {Bellazzini}, {Berihuete}, {Bertone},
  {Bianchi}, {Bienaym{\'e}}, {Blanco-Cuaresma}, {Boch}, {Boeche}, {Bombrun},
  {Borrachero}, {Bossini}, {Bouquillon}, {Bourda}, {Bragaglia}, {Bramante},
  {Breddels}, {Bressan}, {Brouillet}, {Br{\"u}semeister}, {Brugaletta},
  {Bucciarelli}, {Burlacu}, {Busonero}, {Butkevich}, {Buzzi}, {Caffau},
  {Cancelliere}, {Cannizzaro}, {Cantat-Gaudin}, {Carballo}, {Carlucci},
  {Carrasco}, {Casamiquela}, {Castellani}, {Castro-Ginard}, {Charlot},
  {Chemin}, {Chiavassa}, {Cocozza}, {Costigan}, {Cowell}, {Crifo}, {Crosta},
  {Crowley}, {Cuypers}, {Dafonte}, {Damerdji}, {Dapergolas}, {David}, {David},
  {de Laverny}, {De Luise}, {De March}, {de Martino}, {de Souza}, {de Torres},
  {Debosscher}, {del Pozo}, {Delbo}, {Delgado}, {Delgado}, {Di Matteo},
  {Diakite}, {Diener}, {Distefano}, {Dolding}, {Drazinos}, {Dur{\'a}n},
  {Edvardsson}, {Enke}, {Eriksson}, {Esquej}, {Eynard Bontemps}, {Fabre},
  {Fabrizio}, {Faigler}, {Falc{\~a}o}, {Farr{\`a}s Casas}, {Federici},
  {Fedorets}, {Fernique}, {Figueras}, {Filippi}, {Findeisen}, {Fonti},
  {Fraile}, {Fraser}, {Fr{\'e}zouls}, {Gai}, {Galleti}, {Garabato},
  {Garc{\'\i}a-Sedano}, {Garofalo}, {Garralda}, {Gavel}, {Gavras}, {Gerssen},
  {Geyer}, {Giacobbe}, {Gilmore}, {Girona}, {Giuffrida}, {Glass}, {Gomes},
  {Granvik}, {Gueguen}, {Guerrier}, {Guiraud}, {Guti{\'e}rrez-S{\'a}nchez},
  {Haigron}, {Hatzidimitriou}, {Hauser}, {Haywood}, {Heiter}, {Helmi}, {Heu},
  {Hilger}, {Hobbs}, {Hofmann}, {Holland}, {Huckle}, {Hypki}, {Icardi},
  {Jan{\ss}en}, {Jevardat de Fombelle}, {Jonker}, {Juh{\'a}sz}, {Julbe},
  {Karampelas}, {Kewley}, {Klar}, {Kochoska}, {Kohley}, {Kolenberg},
  {Kontizas}, {Kontizas}, {Koposov}, {Kordopatis}, {Kostrzewa-Rutkowska},
  {Koubsky}, {Lambert}, {Lanza}, {Lasne}, {Lavigne}, {Le Fustec}, {Le
  Poncin-Lafitte}, {Lebreton}, {Leccia}, {Leclerc}, {Lecoeur-Taibi},
  {Lenhardt}, {Leroux}, {Liao}, {Licata}, {Lindstr{\o}m}, {Lister}, {Livanou},
  {Lobel}, {L{\'o}pez}, {Managau}, {Mann}, {Mantelet}, {Marchal}, {Marchant},
  {Marconi}, {Marinoni}, {Marschalk{\'o}}, {Marshall}, {Martino}, {Marton},
  {Mary}, {Massari}, {Matijevi{\v{c}}}, {Mazeh}, {McMillan}, {Messina},
  {Michalik}, {Millar}, {Molina}, {Molinaro}, {Moln{\'a}r}, {Montegriffo},
  {Mor}, {Morbidelli}, {Morel}, {Morris}, {Mulone}, {Muraveva}, {Musella},
  {Nelemans}, {Nicastro}, {Noval}, {O'Mullane}, {Ord{\'e}novic},
  {Ord{\'o}{\~n}ez-Blanco}, {Osborne}, {Pagani}, {Pagano}, {Pailler},
  {Palacin}, {Palaversa}, {Panahi}, {Pawlak}, {Piersimoni}, {Pineau}, {Plachy},
  {Plum}, {Poggio}, {Poujoulet}, {Pr{\v{s}}a}, {Pulone}, {Racero}, {Ragaini},
  {Rambaux}, {Ramos-Lerate}, {Regibo}, {Reyl{\'e}}, {Riclet}, {Ripepi}, {Riva},
  {Rivard}, {Rixon}, {Roegiers}, {Roelens}, {Romero-G{\'o}mez}, {Rowell},
  {Royer}, {Ruiz-Dern}, {Sadowski}, {Sagrist{\`a} Sell{\'e}s}, {Sahlmann},
  {Salgado}, {Salguero}, {Sanna}, {Santana-Ros}, {Sarasso}, {Savietto},
  {Schultheis}, {Sciacca}, {Segol}, {Segovia}, {S{\'e}gransan}, {Shih},
  {Siltala}, {Silva}, {Smart}, {Smith}, {Solano}, {Solitro}, {Sordo}, {Soria
  Nieto}, {Souchay}, {Spagna}, {Spoto}, {Stampa}, {Steele},
  {Steidelm{\"u}ller}, {Stephenson}, {Stoev}, {Suess}, {Surdej}, {Szabados},
  {Szegedi-Elek}, {Tapiador}, {Taris}, {Tauran}, {Taylor}, {Teixeira},
  {Terrett}, {Teyssand ier}, {Thuillot}, {Titarenko}, {Torra Clotet}, {Turon},
  {Ulla}, {Utrilla}, {Uzzi}, {Vaillant}, {Valentini}, {Valette}, {van Elteren},
  {Van Hemelryck}, {van Leeuwen}, {Vaschetto}, {Vecchiato}, {Veljanoski},
  {Viala}, {Vicente}, {Vogt}, {von Essen}, {Voss}, {Votruba}, {Voutsinas},
  {Walmsley}, {Weiler}, {Wertz}, {Wevers}, {Wyrzykowski}, {Yoldas},
  {{\v{Z}}erjal}, {Ziaeepour}, {Zorec}, {Zschocke}, {Zucker}, {Zurbach}, \&
  {Zwitter}}]{2018A&A...616A...1G}
{Gaia Collaboration}, {Brown}, A.~G.~A., {Vallenari}, A., {et~al.} 2018, \aap,
  616, A1

\bibitem[{{Giguere} {et~al.}(2016){Giguere}, {Fischer}, {Zhang}, {Matthews},
  {Cameron}, \& {Henry}}]{Giguere16}
{Giguere}, M.~J., {Fischer}, D.~A., {Zhang}, C.~X.~Y., {et~al.} 2016, \apj,
  824, 150

\bibitem[{{Gillett}(1986)}]{1986ASSL..124...61G}
{Gillett}, F.~C. 1986, Astrophysics and Space Science Library, Vol. 124, {IRAS
  observations of cool excess around main sequence stars}, ed. F.~P. {Israel},
  61--69

\bibitem[{{Gizis} {et~al.}(2002){Gizis}, {Reid}, \&
  {Hawley}}]{2002AJ....123.3356G}
{Gizis}, J.~E., {Reid}, I.~N., \& {Hawley}, S.~L. 2002, \aj, 123, 3356

\bibitem[{{Gray}(1984)}]{Gray1984}
{Gray}, D.~F. 1984, \apj, 277, 640

\bibitem[{{Gustafsson} {et~al.}(2008){Gustafsson}, {Edvardsson}, {Eriksson},
  {J{\o}rgensen}, {Nordlund}, \& {Plez}}]{Gustafsson2008}
{Gustafsson}, B., {Edvardsson}, B., {Eriksson}, K., {et~al.} 2008, \aap, 486,
  951

\bibitem[{{Hatzes} {et~al.}(2000){Hatzes}, {Cochran}, {McArthur}, {Baliunas},
  {Walker}, {Campbell}, {Irwin}, {Yang}, {K{\"u}rster}, {Endl}, {Els},
  {Butler}, \& {Marcy}}]{2000ApJ...544L.145H}
{Hatzes}, A.~P., {Cochran}, W.~D., {McArthur}, B., {et~al.} 2000, \apjl, 544,
  L145

\bibitem[{{Haywood} {et~al.}(2014){Haywood}, {Collier Cameron}, {Queloz},
  {Barros}, {Deleuil}, {Fares}, {Gillon}, {Lanza}, {Lovis}, {Moutou}, {Pepe},
  {Pollacco}, {Santerne}, {S{\'e}gransan}, \& {Unruh}}]{2014MNRAS.443.2517H}
{Haywood}, R.~D., {Collier Cameron}, A., {Queloz}, D., {et~al.} 2014, \mnras,
  443, 2517

\bibitem[{{Heiter} {et~al.}(2015){Heiter}, {Jofr{\'e}}, {Gustafsson}, {Korn},
  {Soubiran}, \& {Th{\'e}venin}}]{heiter2015}
{Heiter}, U., {Jofr{\'e}}, P., {Gustafsson}, B., {et~al.} 2015, \aap, 582, A49

\bibitem[{{Hirayama} \& {Moriyama}(1979)}]{1979SoPh...63..251H}
{Hirayama}, T. \& {Moriyama}, F. 1979, \solphys, 63, 251

\bibitem[{{Jeffers} {et~al.}(2017){Jeffers}, {Boro Saikia}, {Barnes}, {Petit},
  {Marsden}, {Jardine}, {Vidotto}, \& {BCool
  Collaboration}}]{2017MNRAS.471L..96J}
{Jeffers}, S.~V., {Boro Saikia}, S., {Barnes}, J.~R., {et~al.} 2017, \mnras,
  471, L96

\bibitem[{{Jeffers} {et~al.}(2014){Jeffers}, {Petit}, {Marsden}, {Morin},
  {Donati}, \& {Folsom}}]{2014A&A...569A..79J}
{Jeffers}, S.~V., {Petit}, P., {Marsden}, S.~C., {et~al.} 2014, \aap, 569, A79

\bibitem[{{Jofr{\'e}} {et~al.}(2014){Jofr{\'e}}, {Heiter}, {Soubiran},
  {Blanco-Cuaresma}, {Worley}, {Pancino}, {Cantat-Gaudin}, {Magrini},
  {Bergemann}, {Gonz{\'a}lez Hern{\'a}ndez}, {Hill}, {Lardo}, {de Laverny},
  {Lind}, {Masseron}, {Montes}, {Mucciarelli}, {Nordlander}, {Recio Blanco},
  {Sobeck}, {Sordo}, {Sousa}, {Tabernero}, {Vallenari}, \& {Van
  Eck}}]{Jofre2014}
{Jofr{\'e}}, P., {Heiter}, U., {Soubiran}, C., {et~al.} 2014, \aap, 564, A133

\bibitem[{{Johns-Krull} {et~al.}(1999){Johns-Krull}, {Valenti}, \&
  {Koresko}}]{JohnsKrull1999}
{Johns-Krull}, C.~M., {Valenti}, J.~A., \& {Koresko}, C. 1999, \apj, 516, 900

\bibitem[{{Johns-Krull} {et~al.}(2004){Johns-Krull}, {Valenti}, \&
  {Saar}}]{JohnsKrull2004}
{Johns-Krull}, C.~M., {Valenti}, J.~A., \& {Saar}, S.~H. 2004, \apj, 617, 1204

\bibitem[{{Keenan} \& {McNeil}(1989)}]{1989ApJS...71..245K}
{Keenan}, P.~C. \& {McNeil}, R.~C. 1989, \apjs, 71, 245

\bibitem[{{Kochukhov} {et~al.}(2010){Kochukhov}, {Makaganiuk}, \&
  {Piskunov}}]{kochukhov10}
{Kochukhov}, O., {Makaganiuk}, V., \& {Piskunov}, N. 2010, \aap, 524, A5

\bibitem[{{Kupka} {et~al.}(1999){Kupka}, {Piskunov}, {Ryabchikova}, {Stempels},
  \& {Weiss}}]{Kupka1999-VALD}
{Kupka}, F., {Piskunov}, N., {Ryabchikova}, T.~A., {Stempels}, H.~C., \&
  {Weiss}, W.~W. 1999, \aaps, 138, 119

\bibitem[{{Kupka} {et~al.}(2000){Kupka}, {Ryabchikova}, {Piskunov}, {Stempels},
  \& {Weiss}}]{Kupka2000-VALD}
{Kupka}, F.~G., {Ryabchikova}, T.~A., {Piskunov}, N.~E., {Stempels}, H.~C., \&
  {Weiss}, W.~W. 2000, Baltic Astronomy, 9, 590

\bibitem[{{Landstreet}(1988)}]{Landstreet1988}
{Landstreet}, J.~D. 1988, \apj, 326, 967

\bibitem[{{Lavail} {et~al.}(2019){Lavail}, {Kochukhov}, \&
  {Hussain}}]{2019A&A...630A..99L}
{Lavail}, A., {Kochukhov}, O., \& {Hussain}, G.~A.~J. 2019, \aap, 630, A99

\bibitem[{{Lavail} {et~al.}(2017){Lavail}, {Kochukhov}, {Hussain}, {Alecian},
  {Herczeg}, \& {Johns-Krull}}]{lavail2017}
{Lavail}, A., {Kochukhov}, O., {Hussain}, G.~A.~J., {et~al.} 2017, \aap, 608,
  A77

\bibitem[{{Lehmann} {et~al.}(2015){Lehmann}, {K{\"u}nstler}, {Carroll}, \&
  {Strassmeier}}]{Lehmann2015}
{Lehmann}, L.~T., {K{\"u}nstler}, A., {Carroll}, T.~A., \& {Strassmeier}, K.~G.
  2015, Astronomische Nachrichten, 336, 258

\bibitem[{{Luck} \& {Heiter}(2005)}]{Luck2005}
{Luck}, R.~E. \& {Heiter}, U. 2005, \aj, 129, 1063

\bibitem[{{Marcy}(1984)}]{Marcy1984}
{Marcy}, G.~W. 1984, \apj, 276, 286

\bibitem[{{Marcy} \& {Basri}(1989)}]{Marcy1989}
{Marcy}, G.~W. \& {Basri}, G. 1989, \apj, 345, 480

\bibitem[{{Marsden} {et~al.}(2014){Marsden}, {Petit}, {Jeffers}, {Morin},
  {Fares}, {Reiners}, {do Nascimento}, {Auri{\`e}re}, {Bouvier}, {Carter},
  {Catala}, {Dintrans}, {Donati}, {Gastine}, {Jardine}, {Konstantinova-Antova},
  {Lanoux}, {Ligni{\`e}res}, {Morgenthaler}, {Ram{\`\i}rez-V{\`e}lez},
  {Th{\'e}ado}, {Van Grootel}, \& {BCool Collaboration}}]{2014MNRAS.444.3517M}
{Marsden}, S.~C., {Petit}, P., {Jeffers}, S.~V., {et~al.} 2014, \mnras, 444,
  3517

\bibitem[{{Mathys} \& {Solanki}(1989)}]{Mathys1989}
{Mathys}, G. \& {Solanki}, S.~K. 1989, \aap, 208, 189

\bibitem[{{Metcalfe} {et~al.}(2013){Metcalfe}, {Buccino}, {Brown}, {Mathur},
  {Soderblom}, {Henry}, {Mauas}, {Petrucci}, {Hall}, \&
  {Basu}}]{2013ApJ...763L..26M}
{Metcalfe}, T.~S., {Buccino}, A.~P., {Brown}, B.~P., {et~al.} 2013, \apjl, 763,
  L26

\bibitem[{{Morin} {et~al.}(2008){Morin}, {Donati}, {Petit}, {Delfosse},
  {Forveille}, {Albert}, {Auri{\`e}re}, {Cabanac}, {Dintrans}, {Fares},
  {Gastine}, {Jardine}, {Ligni{\`e}res}, {Paletou}, {Ramirez Velez}, \&
  {Th{\'e}ado}}]{2008MNRAS.390..567M}
{Morin}, J., {Donati}, J.~F., {Petit}, P., {et~al.} 2008, \mnras, 390, 567

\bibitem[{{Moutou} {et~al.}(2007){Moutou}, {Donati}, {Savalle}, {Hussain},
  {Alecian}, {Bouchy}, {Catala}, {Collier Cameron}, {Udry}, \&
  {Vidal-Madjar}}]{2007A&A...473..651M}
{Moutou}, C., {Donati}, J.~F., {Savalle}, R., {et~al.} 2007, \aap, 473, 651

\bibitem[{{Noyes} {et~al.}(1984){Noyes}, {Hartmann}, {Baliunas}, {Duncan}, \&
  {Vaughan}}]{1984ApJ...279..763N}
{Noyes}, R.~W., {Hartmann}, L.~W., {Baliunas}, S.~L., {Duncan}, D.~K., \&
  {Vaughan}, A.~H. 1984, \apj, 279, 763

\bibitem[{{Petit} {et~al.}(2013){Petit}, {Auri{\`e}re}, {Konstantinova-Antova},
  {Morgenthaler}, {Perrin}, {Roudier}, \& {Donati}}]{2013LNP...857..231P}
{Petit}, P., {Auri{\`e}re}, M., {Konstantinova-Antova}, R., {et~al.} 2013,
  {Magnetic Fields and Convection in the Cool Supergiant Betelgeuse}, ed. J.-P.
  {Rozelot} \& C.~E.~. {Neiner}, Vol. 857, 231

\bibitem[{{Petit} {et~al.}(2002){Petit}, {Donati}, \& {Collier
  Cameron}}]{petit02}
{Petit}, P., {Donati}, J.-F., \& {Collier Cameron}, A. 2002, \mnras, 334, 374

\bibitem[{{Petit} {et~al.}(2014){Petit}, {Louge}, {Th{\'e}ado}, {Paletou},
  {Manset}, {Morin}, {Marsden}, \& {Jeffers}}]{2014PASP..126..469P}
{Petit}, P., {Louge}, T., {Th{\'e}ado}, S., {et~al.} 2014, \pasp, 126, 469

\bibitem[{{Piskunov} {et~al.}(1995){Piskunov}, {Kupka}, {Ryabchikova}, {Weiss},
  \& {Jeffery}}]{Piskunov1995-VALD}
{Piskunov}, N.~E., {Kupka}, F., {Ryabchikova}, T.~A., {Weiss}, W.~W., \&
  {Jeffery}, C.~S. 1995, \aaps, 112, 525

\bibitem[{{Quillen} \& {Thorndike}(2002)}]{2002ApJ...578L.149Q}
{Quillen}, A.~C. \& {Thorndike}, S. 2002, \apjl, 578, L149

\bibitem[{{Rees} \& {Semel}(1979)}]{1979A&A....74....1R}
{Rees}, D.~E. \& {Semel}, M.~D. 1979, \aap, 74, 1

\bibitem[{{Ricker} {et~al.}(2015){Ricker}, {Winn}, {Vanderspek}, {Latham},
  {Bakos}, {Bean}, {Berta-Thompson}, {Brown}, {Buchhave}, {Butler}, {Butler},
  {Chaplin}, {Charbonneau}, {Christensen-Dalsgaard}, {Clampin}, {Deming},
  {Doty}, {De Lee}, {Dressing}, {Dunham}, {Endl}, {Fressin}, {Ge}, {Henning},
  {Holman}, {Howard}, {Ida}, {Jenkins}, {Jernigan}, {Johnson}, {Kaltenegger},
  {Kawai}, {Kjeldsen}, {Laughlin}, {Levine}, {Lin}, {Lissauer}, {MacQueen},
  {Marcy}, {McCullough}, {Morton}, {Narita}, {Paegert}, {Palle}, {Pepe},
  {Pepper}, {Quirrenbach}, {Rinehart}, {Sasselov}, {Sato}, {Seager},
  {Sozzetti}, {Stassun}, {Sullivan}, {Szentgyorgyi}, {Torres}, {Udry}, \&
  {Villasenor}}]{ricker15}
{Ricker}, G.~R., {Winn}, J.~N., {Vanderspek}, R., {et~al.} 2015, Journal of
  Astronomical Telescopes, Instruments, and Systems, 1, 014003

\bibitem[{{Rueedi} {et~al.}(1997){Rueedi}, {Solanki}, {Mathys}, \&
  {Saar}}]{1997A&A...318..429R}
{Rueedi}, I., {Solanki}, S.~K., {Mathys}, G., \& {Saar}, S.~H. 1997, \aap, 318,
  429

\bibitem[{{Ryabchikova} {et~al.}(2015){Ryabchikova}, {Piskunov}, {Kurucz},
  {Stempels}, {Heiter}, {Pakhomov}, \& {Barklem}}]{Ryabchikova2015-VALD}
{Ryabchikova}, T., {Piskunov}, N., {Kurucz}, R.~L., {et~al.} 2015, \physscr,
  90, 054005

\bibitem[{{Ryabchikova} {et~al.}(1997){Ryabchikova}, {Piskunov}, {Kupka}, \&
  {Weiss}}]{Ryabchikova1997-VALD}
{Ryabchikova}, T.~A., {Piskunov}, N.~E., {Kupka}, F., \& {Weiss}, W.~W. 1997,
  Baltic Astronomy, 6, 244

\bibitem[{{Saar}(1988)}]{Saar1988}
{Saar}, S.~H. 1988, \apj, 324, 441

\bibitem[{{See} {et~al.}(2015){See}, {Jardine}, {Vidotto}, {Donati}, {Folsom},
  {Boro Saikia}, {Bouvier}, {Fares}, {Gregory}, {Hussain}, {Jeffers},
  {Marsden}, {Morin}, {Moutou}, {do Nascimento}, {Petit}, {Ros{\'e}n}, \&
  {Waite}}]{2015MNRAS.453.4301S}
{See}, V., {Jardine}, M., {Vidotto}, A.~A., {et~al.} 2015, \mnras, 453, 4301

\bibitem[{{See} {et~al.}(2019){See}, {Matt}, {Folsom}, {Boro Saikia}, {Donati},
  {Fares}, {Finley}, {H{\'e}brard}, {Jardine}, {Jeffers}, {Lehmann}, {Marsden},
  {Mengel}, {Morin}, {Petit}, {Vidotto}, {Waite}, \& {BCool
  Collaboration}}]{2019ApJ...876..118S}
{See}, V., {Matt}, S.~P., {Folsom}, C.~P., {et~al.} 2019, \apj, 876, 118

\bibitem[{{Semel}(1989)}]{semel89}
{Semel}, M. 1989, \aap, 225, 456

\bibitem[{{Semel} {et~al.}(1993){Semel}, {Donati}, \&
  {Rees}}]{1993A&A...278..231S}
{Semel}, M., {Donati}, J.~F., \& {Rees}, D.~E. 1993, \aap, 278, 231

\bibitem[{{Shulyak} {et~al.}(2019){Shulyak}, {Reiners}, {Nagel}, {Tal-Or},
  {Caballero}, {Zechmeister}, {B{\'e}jar}, {Cort{\'e}s-Contreras}, {Martin},
  {Kaminski}, {Ribas}, {Quirrenbach}, {Amado}, {Anglada-Escud{\'e}}, {Bauer},
  {Dreizler}, {Guenther}, {Henning}, {Jeffers}, {K{\"u}rster}, {Lafarga},
  {Montes}, {Morales}, \& {Pedraz}}]{shulyak19}
{Shulyak}, D., {Reiners}, A., {Nagel}, E., {et~al.} 2019, \aap, 626, A86

\bibitem[{{Suresh} {et~al.}(2020){Suresh}, {Chatterjee}, {Cordes}, {Bastian},
  \& {Hallinan}}]{2020arXiv201005929S}
{Suresh}, A., {Chatterjee}, S., {Cordes}, J.~M., {Bastian}, T.~S., \&
  {Hallinan}, G. 2020, arXiv e-prints, arXiv:2010.05929

\bibitem[{{Valenti} \& {Fischer}(2005)}]{2005ApJS..159..141V}
{Valenti}, J.~A. \& {Fischer}, D.~A. 2005, \apjs, 159, 141

\bibitem[{{Valenti} {et~al.}(1995){Valenti}, {Marcy}, \&
  {Basri}}]{1995ApJ...439..939V}
{Valenti}, J.~A., {Marcy}, G.~W., \& {Basri}, G. 1995, \apj, 439, 939

\bibitem[{{Wade} {et~al.}(2001){Wade}, {Bagnulo}, {Kochukhov}, {Land street},
  {Piskunov}, \& {Stift}}]{Wade2001}
{Wade}, G.~A., {Bagnulo}, S., {Kochukhov}, O., {et~al.} 2001, \aap, 374, 265

\bibitem[{{Yu} {et~al.}(2017){Yu}, {Donati}, {H{\'e}brard}, {Moutou}, {Malo},
  {Grankin}, {Hussain}, {Collier Cameron}, {Vidotto}, {Baruteau}, {Alencar},
  {Bouvier}, {Petit}, {Takami}, {Herczeg}, {Gregory}, {Jardine}, {Morin},
  {M{\'e}nard}, \& {Matysse Collaboration}}]{2017MNRAS.467.1342Y}
{Yu}, L., {Donati}, J.~F., {H{\'e}brard}, E.~M., {et~al.} 2017, \mnras, 467,
  1342

\bibitem[{{Zatsarinny} \& {Bartschat}(2006)}]{Zatsarinny2006}
{Zatsarinny}, O. \& {Bartschat}, K. 2006, Journal of Physics B Atomic Molecular
  Physics, 39, 2861

\end{thebibliography}

\appendix

\section{Synthetic view of all activity tracers}

\begin{figure*} 
\centering
\includegraphics[width=16cm]{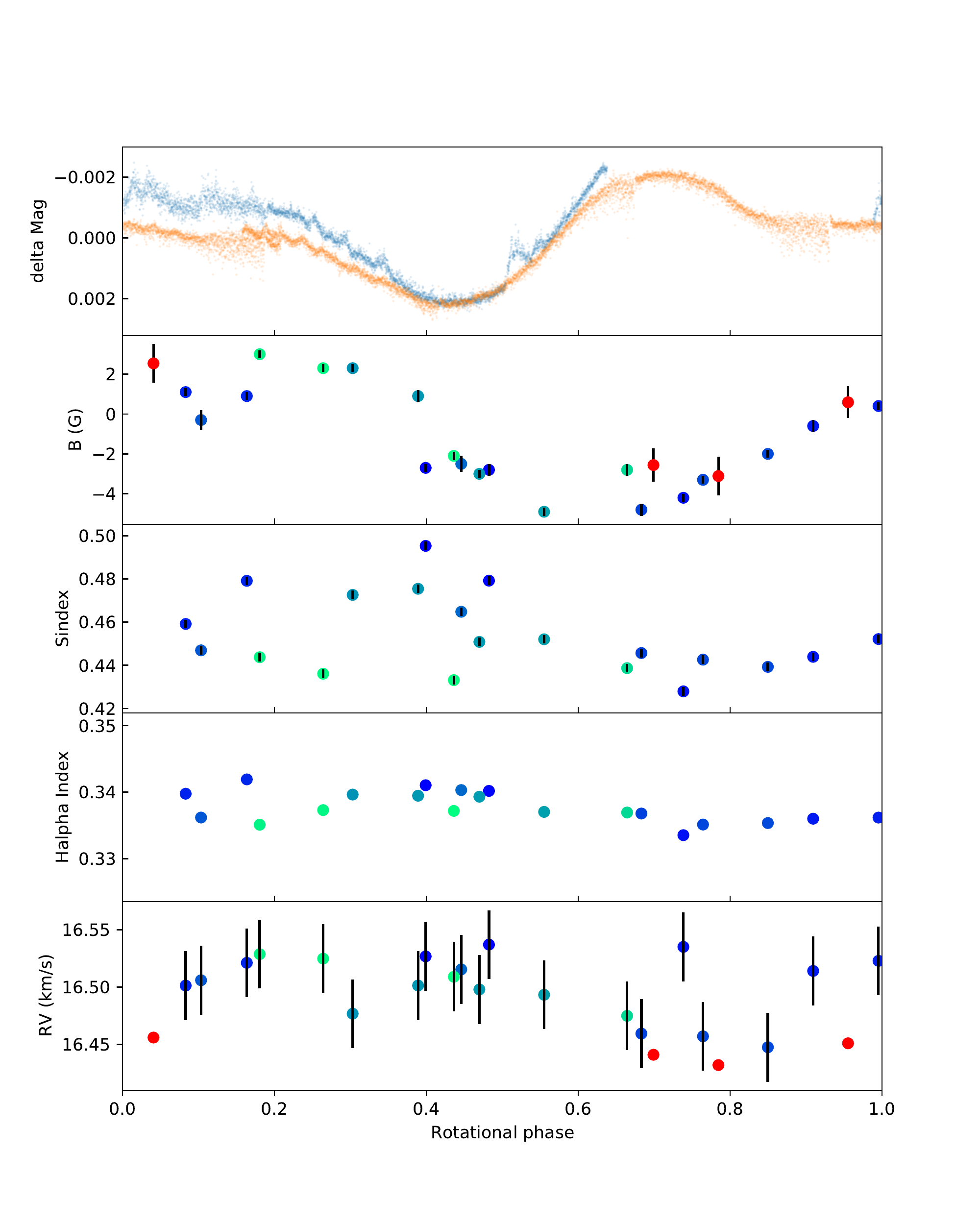} 
\caption{Synthetic view of all activity tracers, as a function of the rotational phase. From top to bottom: normalized TESS light curve, with the first period in blue and the second one in orange. The longitudinal magnetic field, the S index, the H$\alpha$ index, radial velocities.}
\label{fig:all}
\end{figure*}

\section{Extra source of line broadening in the infrared domain}
\label{sec:macro}

\begin{table*}
\centering
\caption{Parameters derived from individual windows in the IR spectrum,  for fits using macroturbulence controlling broadening (upper portion) with \vsin\ inferred from rotation period and radius, and no magnetic field.  Fits using the magnetic field from the Zeeman broadening analysis (and \vsin\ from the visible spectrum analysis) are also presented (bottom portion).  }
\begin{tabular}{lcccccc}
\hline
 & 10500--10920 & 11760--12600 & 15110--15697 & 15815--16390 & 16439--17140 & 21017--22850 \\
\hline
\teff\ (K)    &  5082  & 5051  & 4909  & 4942  & 4983  & 4981  \\ 
\logg         &  4.65  & 4.45  & 4.49  & 4.46  & 4.61  & 4.21  \\ 
\vsin\ (\kms) &  2.59  & 2.59  & 2.59  & 2.59  & 2.59  & 2.59  \\ 
\vmic\ (\kms) &  1.09  & 1.25  & 0.89  & 1.48  & 0.62  & 2.28  \\ 
\vmac\ (\kms) &  2.08  & 2.30  & 3.91  & 4.34  & 3.84  & 4.81  \\ 
metallicity   &  -0.10 & -0.03 & -0.05 & -0.12 & -0.07 & -0.14 \\
\hline
\multicolumn{7}{c}{B=1700 G, f=0.141} \\
\teff\ (K)    & 5066  & 5081  & 4947  & 5048  & 5067  & 5044  \\
\logg         & 4.60  & 4.51  & 4.50  & 4.65  & 4.62  & 4.55  \\
\vsin\ (\kms) & 2.8   & 2.8   & 2.8   & 2.8   & 2.8   & 2.8   \\
\vmic\ (\kms) & 0.72  & 0.85  & 0.51  & 0.89  & 0.0   & 0.92  \\
\vmac\ (\kms) & 1.90  & 1.28  & 3.19  & 3.46  & 3.55  & 4.01  \\
metallicity   & -0.06 & -0.02 & -0.04 & -0.06 & -0.02 & -0.03 \\
\hline
\end{tabular} 
\label{table-params-windows} 
\end{table*}

We find an unexpected trend in macroturbulence towards larger values for longer wavelengths.  Alternatively modelling the line broadening as \vsin\ produces the same trend.  The other best fit parameters do not vary with wavelength, thus this appears to be a feature of the observations, not an error in our methodology.  To investigate if this is real, we fit the solar spectrum obtained with SPIRou for the same spectral windows, assuming \vsin\ $= 2$ \kms\ and using macroturbulence as a free parameter.  We find a trend that is similar but weaker and less consistent, from 2.5 \kms\ in the blue to 3.8 \kms\ in the red.  This is unlikely to be an artefact of the instrument or data reduction process, since calibration images show a consistent line width across the spectrum.  To further check for instrumental effects we fit Gaussian profiles to 77 telluric lines in the observation of \epseri, distributed across the SPIRou domain. While there is some scatter in Gaussian widths, we find consistent widths as a function of wavelength, and widths of the narrower lines are consistent with an R of 70000.  Thus this does not appear to be an instrumental or data reduction effect.  

Rotational broadening (\vsin) should be consistent across a spectrum, apart from a wavelength dependence in limb darkening, which our models account for.  Turbulent broadening may be depth dependent, and widely separated wavelengths have different opacities, thus the physical depth of formation of spectra lines varies with wavelength.  However it is not clear if changes in turbulence with depth could explain a 3.5 \kms\ difference between 6000 \AA\ and 22000 \AA.  The contribution of cool spots to the spectrum increases with wavelength, however it is not clear that cool spots would have much large turbulent broadening than the rest of the photosphere (convection is generally considered to be suppressed in these regions), thus this does not offer an obvious explanation.  

The relative impact of Zeeman broadening increases with wavelength (the Zeeman effect scales as $\lambda^2$ while most other broadening processes scale as $\lambda$), offering a possible explanation.  In Sect \ref{sec:broadening} using Zeeman broadening we find a 1800 G magnetic field covering 14\% of the stellar surface.  We repeated the above spectroscopic analysis assuming this magnetic model, rather than no magnetic field.  This produced a lower reduced $\chi^2$ (by 1 or 2) than for the model with no magnetic field in all windows, except one where reduced $\chi^2$ was largely unchanged.  We find \teff, \logg, and metallicity that are virtually unchanged, a microturbuence that is reduced to $0.78 \pm 0.15$ \kms, however there still is a systematic trend in macroturbulence with wavelength, from 1.90 \kms\ in the blue-most window up to 4.01 \kms\ in the red-most.  Thus including this magnetic field only slightly reduced the wavelength dependence of macroturbulence and is not sufficient to explain it.  

To further investigate whether this wavelength dependent broadening could be Zeeman broadening, we fit the 10500--10920 and 21017--22850 \AA\ windows simultaneously, adding a magnetic field strength and filling factor to the free parameters of \teff, \logg, microturbulence, and metallicity.  Macrotrublence was fixed to the value of the bluer window from above (1.90 \kms), since if left free it tends to a larger intermediate value, too broad for the bluer window.  This produced a best fit magnetic field of 1587 G and filling factor of 0.291, while \teff, \logg, and metallicity were consistent with above (microturbulence was smaller at 0.36 \kms).  
As a second attempt the 11760--12600 and 16439--17140  \AA\ windows were fit simultaneously using the same approach.  This produced a magnetic field of 2107 G and a filling factor of 0.284, and otherwise consistent stellar parameters.
These two models did a good job fitting the line widths in the redder and bluer windows, but both produced filling factors that are inconsistently large for the Zeeman broadening analysis in Sec.\ \ref{sec:broadening}. Specifically, the Fe {\sc i} 15343.79, 15611.14, and 15648.51 \AA\ lines analyzed below all show wings that are too deep with these two models.  Thus including Zeeman broadening is important for accurately determining line broadening in the infrared, but apparently not sufficient to explain the wavelength dependence in line broadening we find here.

We tried allowing the temperature of the magnetic region to differ from the non-magnetic area.  This involves effectively calculating spectra for the two regions using different model atmospheres, interpolated from the same grid and assuming the same \logg, but using different \teff.  The flux ratio between cool to warm regions increases further to the infrared.  If the magnetic regions were cooler, perhaps this could help produce extra Zeeman broadening further into the infrared without increasing the filling actor.  Fitting the 10500--10920 and 21017--22850 \AA\ windows with this model we found a magnetic field of 1652 G, a filling factor of 0.262, a temperature of 4731 K in the magnetic region, and a temperature of 5182 K in the non-magnetic region.  
Fitting the 11760--12600 and 16439--17140 \AA\ windows we found a magnetic field of 2066 G, filling factor of 0.307, temperature in the magnetic region of 4615 K, and temperature in the non-magnetic region of 5321 K.  However, on closer inspection, the temperature in the cooler region appears to be driven largely by the strength of a few lines with very low excitation potentials.  This suggests a possible spot temperature of 4600-4700 K, but does a poor job of providing a model that could explain the wavelength dependent line broadening.

From these tests, it appears that the wavelength dependent line broadening is real, and cannot be fully explained by Zeeman broadening.  This suggests that a depth dependent turbulent velocity should be investigated.  However, that goes beyond the simple micro- macro-turbulence approximation, and may require 3D hydrodynamic model atmospheres to properly investigate. 

\section{Corrections to atomic line data}
\label{sec:linecorr}

  The corrections to atomic line data that we adopted in the SPIRou wavelength range are presented here.  The data were initially extracted from VALD version 3, on 14 Feb.\ 2019, using an `extract stellar' request for the parameters of \epseri, and the default `line list configuration' (i.e.\ selection of input line lists).  An extensive list of empirical corrections to the oscillator strengths were derived by fitting a solar spectrum, as discussed in Sect.~\ref{sec:param_SPIRou}.  The modified line data are presented in Tab.~\ref{table-line-data-corr}. Lines that did not require modification are omitted for brevity.  A number of theoretical transitions were predicted to be detectable but were not present in our observations. These are indicated in Tab.~\ref{table-line-data-corr} by `*'.  A few lines in VALD were apparent duplicates from different sources, specifically components of Mg {\sc i} blends. These are also listed with a `*'.  For some S{\sc i} lines we adopted $\log gf$ from the NIST Atomic Spectra Database, with the original data from \citet{Zatsarinny2006}. Since these provided adequate fits to the observation, these adopted values are included in the Table.   Also included are empirical $\log gf$ corrections for a few Fe {\sc i} lines from \citet{1995ApJ...439..939V}, and Ti {\sc i} lines from \citet{JohnsKrull2004}, since these provided an adequate match to the observations.  

  Additionally, for lines with visible disagreement between the model and observation in the widths of the wings, empirical corrections to the van der Waals damping parameter ($\gamma_6$) were derived.  In some cases $\gamma_6$ values were unavailable, and calculations in the Uns\"old approximation appeared to be insufficient, so empirical values were derived.  The empirical corrections to the transition data likely depend on the parameters of star being investigated and the limitations of the model being employed, thus they should be treated with caution.

While there are a significant number of apparent errors in the oscillator strengths currently available from VALD in the infrared, the list of atomic lines is largely complete. Very few atomic lines in our observations were missing a theoretical counterpart in VALD.  While we did not investigate them in detail, the molecular line list also appears to be largely complete.  However, outside of G and K spectral types, especially towards cooler M dwarfs, the completeness of the line list may become an issue.

\begin{center}

\bigskip 
  
\tablefirsthead{
 \hline
$\lambda$ (\AA) & ion &  $\log gf$ & $\Delta \log gf$ & $\gamma_6$ & $\Delta \gamma_6$ \\
\hline
}
\tablehead{
\multicolumn{6}{l}{Table \ref{table-line-data-corr} continued.}\\
\hline
$\lambda$ (\AA) & ion &  $\log gf$ & $\Delta \log gf$ & $\gamma_6$ & $\Delta \gamma_6$ \\
\hline
}
\tabletail{
  \hline
}
\tablelasttail{
\hline
}
\topcaption{
  Empirical modifications to the line data adopted.
  Lines with a $\log gf$ showing up as `*' were predicted but not detected in the observations (or apparent duplicates).  When a correction to the van der Waals damping parameter ($\gamma_6$) was used, it is indicated.  $\Delta \log gf$ and $\Delta \gamma_6$ are differences with respect to the values from VALD.  Lines without a $\Delta \gamma_6$ value did not have a $\gamma_6$ value available in VALD.
  \label{table-line-data-corr}
}

\begin{supertabular}{llcccc}
10511.588 & P \sc{i}    &  0.10 &  0.23 &  --   &  --  \\
10532.234 & Fe \sc{i}   & -1.67 & -0.19 &  --   &  --  \\
10535.709 & Fe \sc{i}   & -0.14 & -0.03 &  --   &  --  \\
10541.227 & C \sc{i}    & -1.09 &  0.30 &  --   &  --  \\
10555.649 & Fe \sc{i}   & -1.30 & -0.19 &  --   &  --  \\
10577.139 & Fe \sc{i}   & -3.06 &  0.08 &  --   &  --  \\
10582.160 & Si \sc{i}   & -1.04 &  0.13 &  --   &  --  \\
10596.903 & P \sc{i}    &  0.11 &  0.32 &  --   &  --  \\
10602.816 & Si \sc{i}   & -2.17 & -1.23 &  --   &  --  \\
10617.877 & Fe \sc{i}   &  *    &  *    &  --   &  --  \\
10622.592 & Fe \sc{i}   &  *    &  *    &  --   &  --  \\
10627.648 & Si \sc{i}   & -0.30 &  0.56 &  --   &  --  \\
10633.080 & S \sc{i}    &  *    &  *    &  --   &  --  \\
10661.623 & Ti \sc{i}   & -1.88 &  0.04 &  --   &  --  \\
10667.380 & Fe \sc{i}   & -1.90 & -0.44 &  --   &  --  \\
10667.520 & Cr \sc{i}   & -1.80 & -0.32 &  --   &  --  \\
10677.047 & Ti \sc{i}   & -3.00 & -0.48 &  --   &  --  \\
10694.251 & Si \sc{i}   &  0.39 &  0.34 & -7.23 & -0.25\\
10709.942 & Fe \sc{i}   &  *    &  *    &  --   &  --  \\
10717.806 & Fe \sc{i}   & -1.50 & -1.06 &  --   &  --  \\
10726.391 & Ti \sc{i}   & -2.00 &  0.06 &  --   &  --  \\
10727.406 & Si \sc{i}   &  0.50 &  0.28 & -7.07 & -0.11\\
10731.947 & Fe \sc{i}   & -2.30 & -0.46 &  --   &  --  \\
10741.728 & Si \sc{i}   & -0.81 &  0.08 &  --   &  --  \\
10753.004 & Fe \sc{i}   & -2.05 & -0.20 &  --   &  --  \\
10754.281 & Fe \sc{i}   & -1.80 & -0.61 &  --   &  --  \\
10761.445 & Fe \sc{i}   &  *    &  *    &  --   &  --  \\
10768.365 & Al \sc{i}   & -2.10 & -0.60 &  --   &  --  \\
10780.694 & Fe \sc{i}   & -3.60 & -0.31 &  --   &  --  \\
10782.045 & Al \sc{i}   & -1.90 & -0.65 &  --   &  --  \\
10783.050 & Fe \sc{i}   & -2.74 & -0.17 &  --   &  --  \\
10784.562 & Si \sc{i}   & -0.68 &  0.16 &  --   &  --  \\
10786.849 & Si \sc{i}   & -0.10 &  0.20 & -7.56 & -0.29\\
10796.106 & Si \sc{i}   & -1.34 & -0.08 &  --   &  --  \\
10811.053 & Mg \sc{i}   &  0.02 &  0.00 & -6.75 & -0.07\\
10811.084 & Mg \sc{i}   & -0.14 &  0.00 & -6.75 & -0.07\\
10811.097 & Mg \sc{i}   & -1.04 &  0.00 & -6.75 & -0.07\\
10811.122 & Mg \sc{i}   & -1.04 &  0.00 & -6.75 & -0.07\\
10811.158 & Mg \sc{i}   & -0.30 &  0.00 & -6.75 & -0.07\\
10811.198 & Mg \sc{i}   &  *    &  *    &  --   &  --  \\ 
10811.219 & Mg \sc{i}   &  *    &  *    &  --   &  --  \\ 
10818.274 & Fe \sc{i}   & -2.03 & -0.08 &  --   &  --  \\
10825.079 & Ca \sc{i}   &  *    &  *    &  --   &  --  \\
10827.088 & Si \sc{i}   &  0.75 &  0.45 & -8.12 & -0.86\\
10831.938 & Ca \sc{i}   &  *    &  *    &  --   &  --  \\
10832.595 & Ca \sc{i}   &  *    &  *    &  --   &  --  \\
10838.970 & Ca \sc{i}   &  0.07 & -0.17 &  --   &  --  \\
10846.792 & Ca \sc{i}   & -0.31 &  1.01 &  --   &  --  \\
10849.465 & Fe \sc{i}   & -0.66 &  0.78 &  --   &  --  \\
10853.001 & Fe \sc{i}   & -1.22 &  0.23 &  --   &  --  \\
10869.536 & Si \sc{i}   &  0.55 &  0.18 & -7.39 & -0.16\\
10881.758 & Fe \sc{i}   & -3.41 &  0.20 &  --   &  --  \\
10882.809 & Si \sc{i}   & -0.64 &  0.18 &  --   &  --  \\
10884.262 & Fe \sc{i}   & -2.07 & -0.14 &  --   &  --  \\
10885.333 & Si \sc{i}   &  0.15 & -0.08 & -8.13 & -0.80\\
10891.736 & Al \sc{i}   & -0.98 &  0.12 &  --   &  --  \\
10905.710 & Cr \sc{i}   & -0.78 & -0.22 &  --   &  --  \\
10914.244 & Mg \sc{ii}  &  0.05 &  0.03 &  --   &  --  \\
10914.887 & Sr \sc{ii}  & -0.25 &  0.39 &  --   &  --  \\
10915.284 & Mg \sc{ii}  & -1.67 & -0.74 &  --   &  --  \\
11783.265 & Fe \sc{i}   & -2.00 & -0.43 & -7.50 &  0.32\\
11811.558 & Mg \sc{i}   &  *    &  *    &  --   &  --  \\
11820.982 & Mg \sc{i}   &  *    &  *    &  --   &  --  \\
11828.171 & Mg \sc{i}   & -0.18 &  0.15 & -7.27 & -0.08\\
11838.997 & Ca \sc{ii}  &  0.57 &  0.26 &  --   &  --  \\
11848.710 & C \sc{i}    & -0.60 &  0.10 &  --   &  --  \\
11879.580 & C \sc{i}    & -0.48 &  0.13 &  --   &  --  \\
11882.844 & Fe \sc{i}   & -2.44 & -0.78 & -6.99 &  0.83\\
11884.083 & Fe \sc{i}   & -2.79 & -0.71 & -6.78 &  1.04\\
11949.547 & Ti \sc{i}   & -2.28 & -0.71 &  --   &  --  \\
11949.744 & Ca \sc{ii}  & -0.08 & -0.09 & -6.83 &  0.73\\
11973.046 & Fe \sc{i}   & -3.26 & -1.78 & -6.13 &  1.69\\
11973.847 & Ti \sc{i}   & -1.46 & -0.07 &  --   &  --  \\
11984.198 & Si \sc{i}   &  0.30 &  0.06 & -7.40 & -0.10\\
11991.568 & Si \sc{i}   & -0.09 &  0.02 & -7.37 & -0.07\\
12005.397 & Fe \sc{i}   & -1.14 & -0.60 &  --   &  --  \\
12005.547 & Fe \sc{i}   & -0.91 &  0.01 &  --   &  --  \\
12010.578 & Fe \sc{i}   & -1.80 & -0.63 &  --   &  --  \\
12031.504 & Si \sc{i}   &  0.72 &  0.25 & -7.81 & -0.52\\
12053.082 & Fe \sc{i}   & -1.68 & -0.13 &  --   &  --  \\
12081.972 & Si \sc{i}   & -0.61 & -0.09 &  --   &  --  \\
12083.278 & Mg \sc{i}   &  *    &  *    &  --   &  --  \\ 
12083.346 & Mg \sc{i}   & -1.56 & -0.77 &  --   &  --  \\
12083.649 & Mg \sc{i}   &  0.41 &  0.00 & -7.08 & -0.10\\
12084.976 & C \sc{i}    & -0.87 & -0.41 &  --   &  --  \\
12100.181 & Si \sc{i}   & -1.37 & -0.24 & -6.10 &  0.88\\
12105.841 & Ca \sc{i}   & -0.42 & -0.11 & -6.27 &  0.82\\
12110.659 & Si \sc{i}   & -0.59 & -0.46 &  --   &  --  \\
12119.494 & Fe \sc{i}   & -1.86 & -0.23 &  --   &  --  \\
12133.995 & Si \sc{i}   & -1.67 &  0.43 &  --   &  --  \\
12189.241 & Si \sc{i}   & -0.88 &  0.13 &  --   &  --  \\
12190.098 & Fe \sc{i}   & -2.80 & -0.47 &  --   &  --  \\
12227.112 & Fe \sc{i}   & -1.59 & -0.23 &  --   &  --  \\
12342.916 & Fe \sc{i}   & -1.67 & -0.21 &  --   &  --  \\
12390.154 & Si \sc{i}   & -1.85 & -0.08 &  --   &  --  \\
12395.832 & Si \sc{i}   & -1.81 & -0.16 &  --   &  --  \\
12417.936 & Mg \sc{i}   & -1.80 & -0.14 &  --   &  --  \\
12423.029 & Mg \sc{i}   & -1.35 & -0.17 &  --   &  --  \\
12432.273 & K \sc{i}    & -0.31 &  0.13 &  --   &  --  \\
12433.452 & Mg \sc{i}   & -1.06 & -0.09 &  --   &  --  \\
12433.748 & Ca \sc{i}   & -0.39 & -0.32 &  --   &  --  \\
12457.132 & Mg \sc{i}   & -1.59 &  0.67 &  --   &  --  \\
12510.519 & Fe \sc{i}   & -2.01 & -0.40 &  --   &  --  \\
12521.810 & Cr \sc{i}   & -1.57 &  0.01 &  --   &  --  \\
12522.134 & K \sc{i}    &  0.02 &  0.16 &  --   &  --  \\
12556.996 & Fe \sc{i}   & -4.08 & -0.46 &  --   &  --  \\
12569.634 & Co \sc{i}   & -1.29 & -0.30 &  --   &  --  \\
15112.331 & Fe \sc{i}   & -0.85 & -0.23 &  --   &  --  \\
15120.504 & Fe \sc{i}   & -1.52 & -0.72 &  --   &  --  \\
15122.380 & Fe \sc{i}   & -0.49 & -0.28 & -6.92 &  0.59\\
15122.549 & Si \sc{i}   & -1.47 & -0.20 &  --   &  --  \\
15135.306 & Mg \sc{i}   & -1.81 & -0.15 &  --   &  --  \\
15136.124 & Fe \sc{i}   & -0.51 & -0.39 & -6.90 &  0.55\\
15143.089 & Fe \sc{i}   & -1.16 & -0.27 &  --   &  --  \\
15144.051 & Fe \sc{i}   & -0.50 & -0.10 &  --   &  --  \\
15155.208 & Fe \sc{i}   & -1.40 & -0.87 &  --   &  --  \\
15163.067 & K \sc{i}    &  0.51 & -0.18 &  --   &  --  \\
15168.376 & K \sc{i}    &  0.36 & -0.11 &  --   &  --  \\
15176.713 & Fe \sc{i}   & -0.76 & -0.26 & -6.80 &  0.67\\
15182.924 & Fe \sc{i}   & -0.78 & -0.25 &  --   &  --  \\
15183.435 & Fe \sc{i}   & -1.25 & -0.18 &  --   &  --  \\
15201.561 & Fe \sc{i}   & -0.56 & -0.40 &  --   &  --  \\
15207.526 & Fe \sc{i}   & -0.20 & -0.53 & -6.72 &  0.77\\
15213.020 & Fe \sc{i}   & -0.69 & -0.22 &  --   &  --  \\
15219.618 & Fe \sc{i}   & -0.06 &  0.77 & -7.13 &  0.32\\
15231.593 & Mg \sc{i}   & -1.90 & -0.74 &  --   &  --  \\
15231.681 & Mg \sc{i}   & -2.30 & -0.90 &  --   &  --  \\
15231.776 & Mg \sc{i}   & -2.00 & -0.50 &  --   &  --  \\
15239.712 & Fe \sc{i}   & -0.07 & -0.04 &  --   &  --  \\
15243.588 & Si \sc{i}   & -1.14 & -0.27 &  --   &  --  \\
15244.973 & Fe \sc{i}   & -0.08 & -0.01 & -6.96 &  0.49\\
15246.394 & Fe \sc{i}   & -3.06 &  0.55 &  --   &  --  \\
15259.363 & Fe \sc{i}   & -1.60 & -0.35 &  --   &  --  \\
15260.642 & Fe \sc{i}   & -0.70 & -0.23 &  --   &  --  \\
15271.550 & Fe \sc{i}   & -2.04 & -1.29 &  --   &  --  \\
15293.135 & Fe \sc{i}   &  0.01 & -0.13 &  --   &  --  \\
15294.560 & Fe \sc{i}   &  0.38 & -0.34 & -7.02 &  0.47\\
15323.555 & Fe \sc{i}   & -0.71 & -0.12 &  --   &  --  \\
15334.847 & Ti \sc{i}   & -1.10 & -0.14 &  --   &  --  \\
15343.788 & Fe \sc{i}   & -0.67 & -0.09 &  --   &  --  \\
15348.367 & Fe \sc{i}   & -1.66 & -0.56 &  --   &  --  \\
15348.966 & Fe \sc{i}   & -0.86 &  0.40 &  --   &  --  \\
15375.346 & Fe \sc{i}   & -1.39 & -0.40 &  --   &  --  \\
15375.428 & Si \sc{i}   & -1.50 &  0.20 &  --   &  --  \\
15376.831 & Si \sc{i}   & -0.74 & -0.05 & -6.70 &  0.69\\
15381.960 & Fe \sc{i}   & -0.69 & -0.23 &  --   &  --  \\
15394.673 & Fe \sc{i}   & -0.35 & -0.36 & -6.47 &  0.98\\
15398.485 & Fe \sc{i}   & -2.04 & -1.96 &  --   &  --  \\
15400.077 & S \sc{i}    &  0.38 & -0.17 &  --   &  --  \\
15402.331 & S \sc{i}    &  *    &  *    &  --   &  --  \\
15403.724 & S \sc{i}    & -0.35 & -0.67 &  --   &  --  \\
15403.791 & S \sc{i}    &  0.55 &  0.38 &  --   &  --  \\
15405.978 & S \sc{i}    & -1.09 & -0.71 &  --   &  --  \\
15422.261 & S \sc{i}    & -0.35 & -0.71 &  --   &  --  \\
15422.276 & S \sc{i}    &  0.71 &  0.34 &  --   &  --  \\
15427.619 & Fe \sc{i}   & -0.86 & -0.20 &  --   &  --  \\
15444.354 & Fe \sc{i}   &  *    &  *    &  --   &  --  \\
15444.376 & Fe \sc{i}   &  *    &  *    &  --   &  --  \\
15451.298 & Fe \sc{i}   & -0.40 & -0.13 &  --   &  --  \\
15469.816 & S \sc{i}    & -0.20 & -0.15 &  --   &  --  \\
15475.182 & Fe \sc{i}   & -0.70 &  0.71 &  --   &  --  \\
15475.204 & Fe \sc{i}   & -2.10 & -0.09 & -6.50 &  1.04\\
15475.616 & S \sc{i}    & -0.68 & -0.16 &  --   &  --  \\
15475.897 & Fe \sc{i}   & -1.87 & -0.76 &  --   &  --  \\
15476.500 & Fe \sc{i}   & -1.05 & -0.24 &  --   &  --  \\
15478.482 & S \sc{i}    &  0.02 & -0.16 &  --   &  --  \\
15479.603 & Fe \sc{i}   & -0.90 & -0.56 &  --   &  --  \\
15484.334 & Fe \sc{i}   &  *    &  *    &  --   &  --  \\
15485.454 & Fe \sc{i}   & -0.81 &  0.29 &  --   &  --  \\
15486.078 & Fe \sc{i}   &  *    &  *    &  --   &  --  \\
15490.337 & Fe \sc{i}   & -4.90 & -0.32 &  --   &  --  \\
15490.526 & Fe \sc{i}   &  *    &  *    &  --   &  --  \\
15490.881 & Fe \sc{i}   & -0.62 & -0.05 &  --   &  --  \\
15497.000 & Si \sc{i}   & -2.27 & -0.25 &  --   &  --  \\
15497.041 & Fe \sc{i}   & -1.52 & -0.74 &  --   &  --  \\
15500.799 & Fe \sc{i}   & -0.14 & -0.09 &  --   &  --  \\
15501.320 & Fe \sc{i}   &  0.05 & -0.28 &  --   &  --  \\
15506.978 & Si \sc{i}   & -1.63 & -0.34 &  --   &  --  \\
15510.642 & Fe \sc{i}   &  *    &  *    &  --   &  --  \\
15524.308 & Fe \sc{i}   & -1.28 & -0.40 &  --   &  --  \\
15531.751 & Fe \sc{i}   & -0.63 & -0.39 & -6.68 &  0.77\\
15531.802 & Fe \sc{i}   & -0.93 & -0.09 &  --   &  --  \\
15532.449 & Si \sc{i}   & -1.78 & -0.38 &  --   &  --  \\
15534.245 & Fe \sc{i}   & -0.46 & -0.08 & -6.67 &  0.84\\
15537.453 & Fe \sc{i}   & -1.67 & -0.59 &  --   &  --  \\
15537.695 & Fe \sc{i}   & -0.28 & -0.25 &  --   &  --  \\
15542.079 & Fe \sc{i}   & -0.65 & -0.31 & -6.54 &  0.91\\
15543.761 & Ti \sc{i}   & -1.29 & -0.21 &  --   &  --  \\
15547.711 & Fe \sc{i}   & -1.28 & -0.63 &  --   &  --  \\
15550.435 & Fe \sc{i}   & -0.30 & -0.20 &  --   &  --  \\
15551.433 & Fe \sc{i}   & -0.21 &  0.16 &  --   &  --  \\
15557.778 & Si \sc{i}   & -0.78 &  0.03 & -6.97 &  0.42\\
15560.784 & Fe \sc{i}   & -0.36 &  0.11 &  --   &  --  \\
15565.222 & Fe \sc{i}   & -0.77 & -0.21 &  --   &  --  \\
15566.725 & Fe \sc{i}   & -0.37 &  0.31 &  --   &  --  \\
15586.927 & Fe \sc{i}   &  *    &  *    &  --   &  --  \\
15588.259 & Fe \sc{i}   &  0.48 &  0.06 &  --   &  --  \\
15590.046 & Fe \sc{i}   & -0.42 &  0.41 &  --   &  --  \\
15591.490 & Fe \sc{i}   &  0.69 & -0.18 & -6.97 &  0.36\\
15593.749 & Fe \sc{i}   & -1.76 &  0.17 &  --   &  --  \\
15594.396 & Fe \sc{i}   &  *    &  *    &  --   &  --  \\
15598.769 & Fe \sc{i}   & -1.00 & -0.72 &  --   &  --  \\
15598.869 & Fe \sc{i}   & -1.07 & -0.83 &  --   &  --  \\
15602.842 & Ti \sc{i}   & -1.67 & -0.23 &  --   &  --  \\
15604.220 & Fe \sc{i}   &  0.45 & -0.09 &  --   &  --  \\
15605.684 & Ni \sc{i}   & -0.26 & -0.28 &  --   &  --  \\
15611.145 & Fe \sc{i}   & -3.30 &  0.47 &  --   &  --  \\
15613.625 & Fe \sc{i}   & -0.15 &  0.52 &  --   &  --  \\
15617.701 & Fe \sc{i}   &  *    &  *    &  --   &  --  \\
15621.654 & Fe \sc{i}   &  0.01 & -0.58 & -6.61 &  0.84\\
15629.364 & Fe \sc{i}   & -1.82 & -0.81 &  --   &  --  \\
15631.947 & Fe \sc{i}   &  0.10 & -0.02 &  --   &  --  \\
15639.477 & Fe \sc{i}   & -0.87 & -0.80 &  --   &  --  \\
15647.413 & Fe \sc{i}   & -1.08 &  1.21 &  --   &  --  \\
15648.510 & Fe \sc{i}   & -0.63 & -0.03 &  --   &  --  \\
15649.674 & Fe \sc{i}   &  *    &  *    &  --   &  --  \\
15650.563 & Si \sc{i}   &  *    &  *    &  --   &  --  \\
15652.871 & Fe \sc{i}   & -0.04 &  0.12 &  --   &  --  \\
15662.013 & Fe \sc{i}   &  0.12 & -0.25 & -6.92 &  0.53\\
15665.240 & Fe \sc{i}   & -0.41 & -0.08 &  --   &  --  \\
15670.124 & Fe \sc{i}   & -0.80 &  0.18 &  --   &  --  \\
15671.004 & Fe \sc{i}   & -0.45 & -0.23 &  --   &  --  \\
15671.866 & Fe \sc{i}   & -1.19 & -0.13 &  --   &  --  \\
15673.151 & Fe \sc{i}   & -0.58 &  0.15 &  --   &  --  \\
15674.652 & Si \sc{i}   & -0.98 &  0.35 &  --   &  --  \\
15677.012 & Fe \sc{i}   & -0.66 & -0.52 &  --   &  --  \\
15677.519 & Fe \sc{i}   &  0.29 &  0.84 &  --   &  --  \\
15680.060 & Cr \sc{i}   &  0.04 & -0.11 &  --   &  --  \\
15682.016 & Fe \sc{i}   &  *    &  *    &  --   &  --  \\
15682.513 & Fe \sc{i}   & -0.25 &  0.02 &  --   &  --  \\
15683.387 & Fe \sc{i}   & -1.61 &  0.57 &  --   &  --  \\
15686.020 & Fe \sc{i}   & -0.07 & -0.05 &  --   &  --  \\
15686.441 & Fe \sc{i}   &  0.20 & -0.37 &  --   &  --  \\
15687.140 & Fe \sc{i}   & -0.80 & -0.31 &  --   &  --  \\
15691.853 & Fe \sc{i}   &  0.47 & -0.18 & -6.85 &  0.48\\
15693.311 & Mg \sc{i}   & -1.70 & -0.68 &  --   &  --  \\
15693.454 & Mg \sc{i}   & -1.90 & -0.72 &  --   &  --  \\
15693.555 & Mg \sc{i}   & -1.30 &  0.05 &  --   &  --  \\
15816.631 & Fe \sc{i}   & -0.67 & -0.34 &  --   &  --  \\
15822.816 & Fe \sc{i}   & -0.02 & -0.20 &  --   &  --  \\
15827.213 & Si \sc{i}   & -0.75 & -0.10 &  --   &  --  \\
15833.602 & Si \sc{i}   & -0.33 & -0.14 & -7.52 & -0.14\\
15834.164 & Fe \sc{i}   & -0.70 & -0.25 &  --   &  --  \\
15840.190 & Fe \sc{i}   & -0.38 &  0.78 &  --   &  --  \\
15852.580 & C \sc{i}    & -0.44 & -0.18 &  --   &  --  \\
15852.807 & Fe \sc{i}   & -0.84 & -0.29 &  --   &  --  \\
15853.315 & Fe \sc{i}   & -0.76 & -0.20 &  --   &  --  \\
15854.029 & Fe \sc{i}   & -2.69 & -0.68 &  --   &  --  \\
15858.656 & Fe \sc{i}   & -1.33 & -0.85 &  --   &  --  \\
15868.524 & Fe \sc{i}   & -0.03 & -0.12 &  --   &  --  \\
15868.572 & Fe \sc{i}   & -0.09 & -0.52 &  --   &  --  \\
15873.843 & Ti \sc{ii}  & -2.01 & -0.20 &  --   &  --  \\
15878.444 & Fe \sc{i}   & -0.31 &  0.71 &  --   &  --  \\
15884.454 & Si \sc{i}   & -0.91 & -0.08 & -6.83 &  0.56\\
15886.188 & Mg \sc{i}   & -1.70 & -0.18 &  --   &  --  \\
15888.409 & Si \sc{i}   &  0.06 &  0.00 & -7.40 &  0.18\\
15891.160 & Fe \sc{i}   & -0.35 & -0.33 &  --   &  --  \\
15892.395 & Fe \sc{i}   &  0.16 &  0.15 &  --   &  --  \\
15892.769 & Fe \sc{i}   &  0.16 &  0.04 &  --   &  --  \\
15896.555 & Fe \sc{i}   & -0.77 & -1.02 &  --   &  --  \\
15898.016 & Fe \sc{i}   &  0.35 &  0.08 & -6.46 &  0.86\\
15898.890 & Fe \sc{i}   & -1.83 & -1.34 &  --   &  --  \\
15899.252 & Fe \sc{i}   & -0.34 & -0.06 &  --   &  --  \\
15899.586 & Fe \sc{i}   & -1.00 & -0.88 &  --   &  --  \\
15899.710 & Si \sc{i}   & -1.27 & -0.33 &  --   &  --  \\
15901.518 & Fe \sc{i}   & -0.46 &  0.42 &  --   &  --  \\
15906.044 & Fe \sc{i}   & -0.09 & -0.42 &  --   &  --  \\
15909.084 & Fe \sc{i}   & -0.70 & -0.77 &  --   &  --  \\
15909.241 & Fe \sc{i}   & -0.70 & -0.24 &  --   &  --  \\
15912.591 & Fe \sc{i}   & -0.61 & -0.67 &  --   &  --  \\
15912.594 & Mg \sc{i}   & -1.84 & -0.58 &  --   &  --  \\
15913.627 & Fe \sc{i}   & -1.33 & -1.04 &  --   &  --  \\
15914.116 & Si \sc{i}   & -1.73 & -0.46 &  --   &  --  \\
15920.642 & Fe \sc{i}   &  0.38 &  0.01 &  --   &  --  \\
15921.096 & Fe \sc{i}   & -1.36 & -0.44 &  --   &  --  \\
15922.442 & Fe \sc{i}   & -1.10 & -0.49 &  --   &  --  \\
15922.600 & Fe \sc{i}   & -1.00 & -0.48 &  --   &  --  \\
15928.670 & Fe \sc{i}   &  *    &  *    &  --   &  --  \\
15929.472 & Fe \sc{i}   & -0.54 & -0.15 &  --   &  --  \\
15929.843 & Fe \sc{i}   &  *    &  *    &  --   &  --  \\
15932.171 & Fe \sc{i}   &  *    &  *    &  --   &  --  \\
15938.918 & Fe \sc{i}   & -0.10 & -0.17 &  --   &  --  \\
15941.848 & Fe \sc{i}   &  0.03 & -0.24 &  --   &  --  \\
15954.085 & Fe \sc{i}   & -0.58 & -0.08 &  --   &  --  \\
15954.477 & Mg \sc{i}   & -0.95 & -0.20 &  --   &  --  \\
15960.063 & Si \sc{i}   &  0.01 & -0.08 & -6.83 &  0.56\\
15962.558 & Fe \sc{i}   &  0.09 &  0.17 & -6.67 &  0.65\\
15964.865 & Fe \sc{i}   & -0.02 & -0.30 &  --   &  --  \\
15980.725 & Fe \sc{i}   &  0.70 & -0.26 & -6.77 &  0.55\\
15982.072 & Fe \sc{i}   & -0.39 &  0.37 &  --   &  --  \\
16150.762 & Ca \sc{i}   & -0.24 & -0.21 &  --   &  --  \\
16152.714 & Ni \sc{i}   & -1.74 & -0.28 &  --   &  --  \\
16155.236 & Ca \sc{i}   & -0.70 & -0.21 &  --   &  --  \\
16156.557 & Fe \sc{i}   & -0.40 & -0.10 &  --   &  --  \\
16157.364 & Ca \sc{i}   & -0.18 & -0.43 &  --   &  --  \\
16163.691 & Si \sc{i}   & -0.93 & -0.07 & -7.09 &  0.35\\
16165.029 & Fe \sc{i}   &  0.64 & -0.35 & -6.73 &  0.59\\
16174.975 & Fe \sc{i}   &  0.16 & -0.03 & -6.86 &  0.46\\
16186.475 & Si \sc{i}   & -1.28 &  0.34 &  --   &  --  \\
16195.060 & Fe \sc{i}   &  0.14 & -0.33 &  --   &  --  \\
16197.075 & Ca \sc{i}   &  0.10 & -0.15 & -6.55 &  0.70\\
16201.513 & Fe \sc{i}   & -0.51 & -0.18 &  --   &  --  \\
16202.330 & Fe \sc{i}   &  *    &  *    &  --   &  --  \\
16203.328 & Fe \sc{i}   & -0.68 &  0.33 &  --   &  --  \\
16204.252 & Fe \sc{i}   &  0.06 & -0.15 &  --   &  --  \\
16213.537 & Fe \sc{i}   &  0.24 & -0.08 &  --   &  --  \\
16215.670 & Si \sc{i}   & -0.81 & -0.18 & -6.89 &  0.55\\
16225.618 & Fe \sc{i}   &  0.09 & -0.20 &  --   &  --  \\
16227.151 & Fe \sc{i}   & -0.84 &  1.08 &  --   &  --  \\
16232.518 & Fe \sc{i}   &  *    &  *    &  --   &  --  \\
16235.966 & Fe \sc{i}   & -0.23 & -0.21 &  --   &  --  \\
16238.952 & Fe \sc{i}   &  *    &  *    &  --   &  --  \\
16240.870 & Fe \sc{i}   & -0.74 & -0.52 &  --   &  --  \\
16241.833 & Si \sc{i}   & -0.74 &  0.03 & -7.33 &  0.11\\
16245.763 & Fe \sc{i}   & -0.73 &  0.94 &  --   &  --  \\
16246.460 & Fe \sc{i}   & -0.12 & -0.18 &  --   &  --  \\
16252.550 & Fe \sc{i}   & -0.42 & -0.11 &  --   &  --  \\
16258.912 & Fe \sc{i}   & -0.81 &  0.22 &  --   &  --  \\
16272.468 & Fe \sc{i}   & -0.75 &  0.51 &  --   &  --  \\
16284.769 & Fe \sc{i}   &  0.15 &  0.32 &  --   &  --  \\
16292.840 & Fe \sc{i}   & -0.48 & -0.32 &  --   &  --  \\
16310.501 & Ni \sc{i}   &  0.06 & -0.16 &  --   &  --  \\
16316.320 & Fe \sc{i}   &  0.74 & -0.33 & -6.72 &  0.60\\
16324.451 & Fe \sc{i}   & -0.56 &  0.00 & -7.11 &  0.38\\
16331.524 & Fe \sc{i}   & -0.58 & -0.18 &  --   &  --  \\
16333.141 & Fe \sc{i}   & -1.44 & -0.80 &  --   &  --  \\
16333.928 & C \sc{i}    & -1.40 & -0.27 &  --   &  --  \\
16346.857 & Si \sc{i}   & -0.70 &  0.39 &  --   &  --  \\
16363.103 & Ni \sc{i}   &  0.62 &  0.03 & -8.13 & -0.65\\
16364.748 & Mg \sc{i}   & -1.30 & -0.46 &  --   &  --  \\
16364.850 & Mg \sc{i}   & -1.30 & -0.30 &  --   &  --  \\
16364.960 & Mg \sc{i}   & -0.68 &  0.49 & -5.42 &  none\\
16366.337 & Fe \sc{i}   & -0.36 &  1.00 &  --   &  --  \\
16377.388 & Fe \sc{i}   & -0.23 &  0.24 &  --   &  --  \\
16380.176 & Si \sc{i}   & -0.70 & -0.23 &  --   &  --  \\
16381.204 & Fe \sc{i}   & -0.14 &  0.11 &  --   &  --  \\
16381.534 & Si \sc{i}   & -0.50 & -0.04 &  --   &  --  \\
16381.814 & Fe \sc{i}   & -0.90 & -0.88 &  --   &  --  \\
16382.251 & Fe \sc{i}   &  0.35 & -0.04 & -7.00 &  0.33\\
16384.141 & Fe \sc{i}   & -0.19 &  0.55 &  --   &  --  \\
16412.982 & Si \sc{i}   & -1.60 & -0.87 &  --   &  --  \\
16444.816 & Fe \sc{i}   &  0.18 & -0.48 & -6.74 &  0.71\\
16466.921 & Fe \sc{i}   &  0.08 &  0.08 &  --   &  --  \\
16468.533 & C \sc{i}    & -1.08 &  0.07 &  --   &  --  \\
16471.753 & Fe \sc{i}   & -0.68 & -0.71 &  --   &  --  \\
16474.077 & Fe \sc{i}   & -0.42 &  0.54 &  --   &  --  \\
16476.933 & Fe \sc{i}   & -0.47 &  0.12 &  --   &  --  \\
16481.228 & Fe \sc{i}   & -0.30 & -0.14 &  --   &  --  \\
16486.666 & Fe \sc{i}   &  0.24 & -0.54 & -6.62 &  0.83\\
16489.788 & Fe \sc{i}   &  *    &  *    &  --   &  --  \\
16489.987 & Mn \sc{i}   &  *    &  *    &  --   &  --  \\
16494.427 & Fe \sc{i}   & -0.80 & -0.28 &  --   &  --  \\
16494.500 & Fe \sc{i}   & -0.84 & -0.38 &  --   &  --  \\
16494.702 & Fe \sc{i}   & -1.10 & -0.47 &  --   &  --  \\
16504.140 & Fe \sc{i}   &  *    &  *    &  --   &  --  \\
16506.293 & Fe \sc{i}   & -0.37 &  0.09 &  --   &  --  \\
16517.223 & Fe \sc{i}   &  0.48 & -0.20 & -7.03 &  0.29\\
16518.940 & Fe \sc{i}   &  *    &  *    &  --   &  --  \\
16519.147 & Fe \sc{i}   &  *    &  *    &  --   &  --  \\
16522.074 & Fe \sc{i}   & -0.05 & -0.38 &  --   &  --  \\
16524.466 & Fe \sc{i}   &  0.45 & -0.24 & -6.70 &  0.62\\
16531.983 & Fe \sc{i}   & -0.05 &  0.77 &  --   &  --  \\
16537.994 & Fe \sc{i}   & -0.42 &  0.45 &  --   &  --  \\
16539.193 & Fe \sc{i}   & -0.10 &  0.02 & -7.00 &  0.33\\
16540.870 & Fe \sc{i}   & -0.62 &  0.05 &  --   &  --  \\
16541.423 & Fe \sc{i}   & -0.39 &  1.31 &  --   &  --  \\
16541.962 & Fe \sc{i}   & -0.36 & -0.51 &  --   &  --  \\
16542.660 & S \sc{i}    & -0.39 & -0.32 &  --   &  --  \\
16544.667 & Fe \sc{i}   & -0.32 & -0.29 &  --   &  --  \\
16550.383 & Ni \sc{i}   &  0.33 &  0.06 &  --   &  --  \\
16551.994 & Fe \sc{i}   &  0.15 & -0.19 &  --   &  --  \\
16556.347 & Si \sc{i}   & -1.00 & -0.05 &  --   &  --  \\
16556.484 & Fe \sc{i}   &  *    &  *    &  --   &  --  \\
16556.674 & Fe \sc{i}   &  *    &  *    &  --   &  --  \\
16557.148 & Fe \sc{i}   & -0.37 &  0.71 &  --   &  --  \\
16559.677 & Fe \sc{i}   & -0.26 & -0.47 &  --   &  --  \\
16570.510 & Fe \sc{i}   &  *    &  *    &  --   &  --  \\
16575.271 & Fe \sc{i}   & -2.35 & -0.86 &  --   &  --  \\
16578.064 & Fe \sc{i}   & -2.77 & -0.76 &  --   &  --  \\
16581.383 & Fe \sc{i}   &  *    &  *    &  --   &  --  \\
16584.480 & Ni \sc{i}   & -0.38 &  0.33 &  --   &  --  \\
16586.051 & Fe \sc{i}   & -1.31 & -0.56 &  --   &  --  \\
16587.493 & Fe \sc{i}   & -1.03 & -0.29 &  --   &  --  \\
16589.439 & Ni \sc{i}   & -0.57 & -0.23 &  --   &  --  \\
16607.634 & Fe \sc{i}   & -0.57 &  1.03 &  --   &  --  \\
16612.761 & Fe \sc{i}   &  0.09 & -0.20 &  --   &  --  \\
16619.737 & Fe \sc{i}   & -1.43 & -0.57 &  --   &  --  \\
16624.881 & Mg \sc{i}   & -1.78 & -0.41 &  --   &  --  \\
16629.836 & Fe \sc{i}   &  *    &  *    &  --   &  --  \\
16632.019 & Mg \sc{i}   &  *    &  *    &  --   &  --  \\ 
16632.230 & Mg \sc{i}   & -1.19 & -0.09 &  --   &  --  \\
16632.503 & Fe \sc{i}   & -1.32 & -1.08 &  --   &  --  \\
16640.640 & Fe \sc{i}   &  *    &  *    &  --   &  --  \\
16647.246 & Fe \sc{i}   &  *    &  *    &  --   &  --  \\
16648.203 & Fe \sc{i}   & -0.29 &  0.20 &  --   &  --  \\
16649.877 & Ca \sc{ii}  &  0.72 &  0.08 &  --   &  --  \\
16652.387 & Fe \sc{i}   & -0.80 & -0.37 &  --   &  --  \\
16652.798 & Fe \sc{i}   & -0.60 & -0.12 &  --   &  --  \\
16661.379 & Fe \sc{i}   &  0.22 &  0.97 &  --   &  --  \\
16666.773 & Fe \sc{i}   & -0.87 & -0.22 &  --   &  --  \\
16673.706 & Ni \sc{i}   &  0.14 & -0.25 &  --   &  --  \\
16680.770 & Si \sc{i}   & -0.10 &  0.04 & -7.20 &  0.24\\
16693.072 & Fe \sc{i}   & -0.28 & -0.14 &  --   &  --  \\
16711.282 & Fe \sc{i}   &  *    &  *    &  --   &  --  \\
16718.957 & Al \sc{i}   &  0.05 & -0.10 & -7.09 &  none\\
16721.462 & Fe \sc{i}   & -0.39 &  0.19 &  --   &  --  \\
16724.685 & Fe \sc{i}   & -0.73 & -0.32 &  --   &  --  \\
16725.440 & Fe \sc{i}   & -0.90 & -0.34 &  --   &  --  \\
16728.309 & Fe \sc{i}   & -1.40 & -0.65 &  --   &  --  \\
16729.672 & Si \sc{i}   &  *    &  *    &  --   &  --  \\
16737.240 & Fe \sc{i}   &  *    &  *    &  --   &  --  \\
16739.311 & Fe \sc{i}   & -1.17 & -0.37 &  --   &  --  \\
16750.564 & Al \sc{i}   &  0.38 & -0.03 & -7.16 &  none\\
16753.065 & Fe \sc{i}   &  0.31 & -0.09 &  --   &  --  \\
16757.642 & Co \sc{i}   & -1.50 & -0.58 &  --   &  --  \\
16760.218 & Mg \sc{ii}  &  0.47 & -0.01 &  --   &  --  \\
16763.360 & Al \sc{i}   & -0.43 &  0.12 &  --   &  --  \\
16828.159 & Si \sc{i}   & -1.05 & -0.02 & -7.21 &  0.23\\
16833.052 & Fe \sc{i}   & -1.02 & -0.13 &  --   &  --  \\
16843.228 & Fe \sc{i}   & -1.22 &  0.10 &  --   &  --  \\
16843.877 & Fe \sc{i}   & -1.90 & -1.27 &  --   &  --  \\
16853.089 & Ni \sc{i}   &  *    &  *    &  --   &  --  \\
16853.467 & Fe \sc{i}   &  *    &  *    &  --   &  --  \\
16854.936 & C \sc{i}    & -0.93 & -0.12 &  --   &  --  \\
16856.442 & Fe \sc{i}   & -0.93 & -0.84 &  --   &  --  \\
16857.135 & Fe \sc{i}   & -1.75 & -1.50 &  --   &  --  \\
16858.523 & Fe \sc{i}   & -1.54 & -1.68 &  --   &  --  \\
16864.079 & Fe \sc{i}   & -0.86 &  0.68 &  --   &  --  \\
16865.513 & Fe \sc{i}   & -0.87 & -0.12 &  --   &  --  \\
16869.950 & Fe \sc{i}   & -0.74 & -0.32 &  --   &  --  \\
16874.116 & Fe \sc{i}   & -0.80 & -0.64 &  --   &  --  \\
16878.540 & Fe \sc{i}   & -1.33 & -0.55 &  --   &  --  \\
16883.606 & Fe \sc{i}   & -1.41 & -0.47 &  --   &  --  \\
16884.809 & Fe \sc{i}   & -1.07 &  0.65 &  --   &  --  \\
16889.473 & Fe \sc{i}   & -1.31 & -0.47 &  --   &  --  \\
16890.380 & C \sc{i}    &  0.32 & -0.25 & -6.90 &  0.61\\
16892.384 & Fe \sc{i}   & -0.61 &  0.19 &  --   &  --  \\
16893.954 & Fe \sc{i}   & -1.01 &  0.47 &  --   &  --  \\
16898.883 & Fe \sc{i}   & -0.72 &  0.19 &  --   &  --  \\
16900.231 & Fe \sc{i}   & -0.94 & -0.13 &  --   &  --  \\
16910.683 & Fe \sc{i}   &  *    &  *    &  --   &  --  \\
16927.611 & Fe \sc{i}   & -0.31 & -0.15 &  --   &  --  \\
16928.623 & Fe \sc{i}   & -0.73 &  0.33 &  --   &  --  \\
16930.962 & Fe \sc{i}   & -1.12 & -0.19 &  --   &  --  \\
16947.417 & Ca \sc{i}   &  *    &  *    &  --   &  --  \\
16954.102 & C \sc{i}    & -0.71 &  0.25 &  --   &  --  \\
16957.794 & Si \sc{i}   & -1.20 & -0.13 &  --   &  --  \\
16996.265 & Ni \sc{i}   &  0.44 & -0.02 & -7.56 & -0.09\\
17001.025 & Ni \sc{i}   &  0.34 & -0.04 & -6.93 &  0.57\\
17004.631 & Fe \sc{i}   &  *    &  *    &  --   &  --  \\
17007.489 & Fe \sc{i}   & -0.94 & -0.37 &  --   &  --  \\
17012.728 & Fe \sc{i}   &  *    &  *    &  --   &  --  \\
17018.033 & Ti \sc{i}   &  0.50 & -0.26 &  --   &  --  \\
17018.624 & Fe \sc{i}   & -1.30 & -0.33 &  --   &  --  \\
17025.116 & Fe \sc{i}   & -1.31 & -0.94 &  --   &  --  \\
17032.896 & Fe \sc{i}   & -1.36 & -0.43 &  --   &  --  \\
17033.659 & Fe \sc{i}   & -0.72 & -0.47 &  --   &  --  \\
17037.787 & Fe \sc{i}   & -0.28 &  0.58 &  --   &  --  \\
17040.099 & Fe \sc{i}   &  *    &  *    &  --   &  --  \\
17047.651 & Fe \sc{i}   &  *    &  *    &  --   &  --  \\
17052.181 & Fe \sc{i}   & -0.53 &  0.21 &  --   &  --  \\
17052.876 & Fe \sc{i}   &  *    &  *    &  --   &  --  \\
17061.249 & Fe \sc{i}   & -0.93 &  0.36 &  --   &  --  \\
17064.887 & Fe \sc{i}   & -0.39 & -0.16 &  --   &  --  \\
17065.265 & Fe \sc{i}   & -0.69 & -0.19 &  --   &  --  \\
17067.529 & Fe \sc{i}   & -0.03 &  1.36 &  --   &  --  \\
17070.548 & Fe \sc{i}   & -0.83 & -0.16 &  --   &  --  \\
17072.825 & Fe \sc{i}   & -2.19 & -0.28 &  --   &  --  \\
17075.120 & Fe \sc{i}   & -0.92 &  1.20 &  --   &  --  \\
17085.630 & Mg \sc{i}   & -1.69 &  0.22 &  --   &  --  \\
17086.250 & C \sc{i}    & -1.54 &  0.34 &  --   &  --  \\
17091.304 & Fe \sc{i}   &  *    &  *    &  --   &  --  \\
17094.434 & Fe \sc{i}   & -0.60 & -0.23 &  --   &  --  \\
17108.631 & Mg \sc{i}   & -0.17 & -0.24 & -6.91 &  none\\
17112.447 & P \sc{i}    &  1.03 &  0.53 &  --   &  --  \\
17115.719 & Fe \sc{i}   & -1.07 &  0.46 &  --   &  --  \\
17120.503 & Ni \sc{i}   & -0.42 & -0.22 &  --   &  --  \\
17130.952 & Fe \sc{i}   & -0.45 &  0.45 &  --   &  --  \\
17132.928 & Fe \sc{i}   & -0.73 &  0.36 &  --   &  --  \\
17134.200 & Fe \sc{i}   & -1.01 &  0.28 &  --   &  --  \\
17137.105 & Fe \sc{i}   & -0.83 & -0.67 &  --   &  --  \\
17138.897 & Fe \sc{i}   & -0.88 &  0.66 &  --   &  --  \\
17151.667 & Fe \sc{i}   & -1.07 & -0.70 &  --   &  --  \\
17232.229 & Fe \sc{i}   & -0.88 &  0.54 &  --   &  --  \\
17233.171 & Fe \sc{i}   & -0.79 &  0.84 &  --   &  --  \\
21047.143 & Si \sc{i}   & -0.33 &  0.13 &  --   &  --  \\
21056.253 & Si \sc{i}   & -0.47 & -0.11 &  --   &  --  \\
21060.711 & Mg \sc{i}   & -0.45 &  0.04 & -5.97 &  none\\
21060.891 & Mg \sc{i}   & -0.37 &  0.00 & -6.00 &  none\\
21061.091 & Mg \sc{i}   & -0.29 & -0.04 & -5.76 &  none\\
21093.029 & Al \sc{i}   & -0.48 & -0.17 & -6.96 &  none\\
21139.759 & Si \sc{i}   & -0.47 & -0.03 &  --   &  --  \\
21144.154 & Si \sc{i}   & -0.68 & -0.11 &  --   &  --  \\
21162.035 & Fe \sc{i}   & -0.31 &  0.03 &  --   &  --  \\
21163.755 & Al \sc{i}   &  0.11 &  0.12 & -7.05 &  none\\
21204.829 & Si \sc{i}   & -0.15 &  0.18 &  --   &  --  \\
21208.141 & Mg \sc{i}   & -1.00 &  1.04 &  --   &  --  \\
21211.510 & C \sc{i}    &  0.13 &  0.20 &  --   &  --  \\
21233.658 & Mg \sc{i}   &  *    &  *    &  --   &  --  \\
21238.466 & Fe \sc{i}   & -1.50 & -0.08 &  --   &  --  \\
21246.817 & Ca \sc{i}   &  *    &  *    &  --   &  --  \\
21259.897 & C \sc{i}    &  0.59 &  0.09 & -6.48 &  0.85\\
21354.198 & Si \sc{i}   &  0.13 & -0.06 & -7.34 &  0.04\\
21425.785 & Si \sc{i}   &  0.08 &  0.30 &  --   &  --  \\
21441.924 & Si \sc{i}   &  0.15 &  0.45 &  --   &  --  \\
21489.572 & Fe \sc{i}   &  *    &  *    &  --   &  --  \\
21523.114 & Si \sc{i}   &  *    &  *    &  --   &  --  \\
21756.947 & Fe \sc{i}   & -0.70 &  0.15 &  --   &  --  \\
21761.020 & Mg \sc{i}   &  *    &  *    &  --   &  --  \\
21779.660 & Si \sc{i}   &  0.02 & -0.39 & -6.86 &  0.47\\
21782.944 & Ti \sc{i}   & -1.18 & -0.01 &  --   &  --  \\
21816.566 & Fe \sc{i}   & -0.20 &  0.25 &  --   &  --  \\
21820.661 & Fe \sc{i}   & -0.92 & -1.05 &  --   &  --  \\
21851.381 & Fe \sc{i}   & -3.76 & -0.15 &  --   &  --  \\
21858.065 & Fe \sc{i}   &  0.10 &  0.40 &  --   &  --  \\
21879.324 & Si \sc{i}   &  0.18 & -0.23 & -6.68 &  0.65\\
21880.871 & Fe \sc{i}   &  *    &  *    &  --   &  --  \\
21882.987 & Fe \sc{i}   & -0.04 & -0.18 &  --   &  --  \\
21894.983 & Fe \sc{i}   & -0.04 &  0.31 &  --   &  --  \\
21897.391 & Ti \sc{i}   & -1.30 &  0.17 &  --   &  --  \\
21951.294 & Fe \sc{i}   &  *    &  *    &  --   &  --  \\
22056.400 & Na \sc{i}   &  0.27 & -0.02 & -7.13 &  none\\
22072.550 & Si \sc{i}   & -1.15 & -0.21 &  --   &  --  \\
22083.662 & Na \sc{i}   & -0.02 & -0.01 & -7.09 &  none\\
22139.693 & Fe \sc{i}   &  *    &  *    &  --   &  --  \\
22178.155 & Fe \sc{i}   &  *    &  *    &  --   &  --  \\
22211.238 & Ti \sc{i}   & -1.64 &  0.14 &  --   &  --  \\
22232.858 & Ti \sc{i}   & -1.56 &  0.13 &  --   &  --  \\
22260.179 & Fe \sc{i}   & -0.99 & -0.05 &  --   &  --  \\
22274.022 & Ti \sc{i}   & -1.63 &  0.17 &  --   &  --  \\
22310.617 & Ti \sc{i}   & -1.94 &  0.13 &  --   &  --  \\
22419.976 & Fe \sc{i}   & -0.22 & -0.07 &  --   &  --  \\
22493.671 & Fe \sc{i}   & -1.14 & -0.25 &  --   &  --  \\
22537.534 & Si \sc{i}   & -0.31 & -0.08 & -6.66 &  0.60\\
22563.828 & S \sc{i}    & -0.00 &  0.26 &  --   &  --  \\
22609.238 & Fe \sc{i}   & -1.33 & -0.64 &  --   &  --  \\
22619.838 & Fe \sc{i}   & -0.65 & -0.28 & -6.51 &  1.03\\
22626.723 & Ca \sc{i}   & -0.38 & -0.16 &  --   &  --  \\
22651.177 & Ca \sc{i}   &  0.55 & -0.30 & -6.73 &  0.60\\
22665.757 & Si \sc{i}   & -0.26 &  0.42 &  --   &  --  \\
22707.738 & S \sc{i}    &  0.21 & -0.23 &  --   &  --  \\
22807.745 & Mg \sc{i}   & -0.32 & -0.34 & -6.04 &  none\\
22807.775 & Mg \sc{i}   & -1.25 & -0.21 & -6.30 &  none\\
22807.775 & Mg \sc{i}   & -1.25 & -0.20 & -6.30 &  none\\
22808.025 & Mg \sc{i}   & -0.29 & -0.14 & -6.48 &  none\\
22808.033 & Mg \sc{i}   &  *    &  *    &  --   &  --  \\ 
22808.265 & Mg \sc{i}   & -0.50 & -0.19 & -6.30 &  none\\
22812.586 & Fe \sc{i}   & -1.50 & -0.60 &  --   &  --  \\
22832.364 & Fe \sc{i}   & -1.34 & -0.07 &  --   &  --  \\

\end{supertabular}

\end{center}

\end{document}